\documentclass[twocolumn,aps,multicol,amsmath,amssymb]{revtex4-1}

\usepackage{graphicx}
\usepackage{epstopdf}
\usepackage{bm}
\usepackage{subfigure}
\usepackage{sidecap}
\graphicspath{{Figures/}}
\usepackage{times}

\newcommand{\al}{\alpha}
\newcommand{\bt}{\beta}

\newcommand{\la}{\lambda}

\newcommand{\vare}{\varepsilon}
\newcommand{\ze}{\zeta}

\newcommand{\bsig}{{\bm\sigma}}

\newcommand{\be}{\begin{equation}}
\newcommand{\ee}{\end{equation}}

\newcommand{\bea}{\begin{eqnarray}}
\newcommand{\eea}{\end{eqnarray}}
\newcommand{\bd}{\begin{displaymath}}
\newcommand{\ed}{\end{displaymath}}
\newcommand{\ba}{\begin{array}}
\newcommand{\ea}{\end{array}}
\newcommand{\bi}{\begin{itemize}}
\newcommand{\ei}{\end{itemize}}
\newcommand{\bc}{\begin{center}}
\newcommand{\ec}{\end{center}}
\newcommand{\bfl}{\begin{flushleft}}
\newcommand{\efl}{\end{flushleft}}
\newcommand{\bfr}{\begin{flushright}}
\newcommand{\efr}{\end{flushright}}
\newcommand{\non}{\nonumber}

\newcommand{\bl}{\begin{aligned}}
\newcommand{\el}{\end{aligned}}

\newcommand{\hatt}{\hat{t}}
\newcommand{\hh}{\hat{h}}

\newcommand{\hG}{\hat{G}}
\newcommand{\hV}{\hat{V}}

\newcommand{\hE}{\hat{E}}

\newcommand{\ua}{\uparrow}
\newcommand{\da}{\downarrow}

\newcommand{\tL}{\tilde{\Lambda}}

\newcommand{\tbq}{\tilde{{\bf q}}}
\newcommand{\tV}{\tilde{V}}

\newcommand{\om}{i\omega_n}

\newcommand{\fs}{\frac{1}{2}}

\def\ket#1{\left\vert #1 \right\rangle}
\def\dg{^{\dagger}}

\def\br{{\bf r}}\def\bR{{\bf R}}
\def\bk{{\bf k}} \def\bq{{\bf q}} 
  \def\bb{{\bf b}}
\def\bg{{\bf g}} 
  
 \def\bd{{\bf d}}  
\def\bB{{\bf B}}  
\def\hbx{\hat{{\bf x}}} \def\hby{\hat{{\bf y}}} 

\def\da{\downarrow} \def\ua{\uparrow}
 
\def\dg{\dagger}

\def\bra{\langle}
\def\ket{\rangle}
\def\={\!\!\!&=&\!\!\!}
\def\+{\!\!\!&&\!\!\!+~}
\def\-{\!\!\!&&\!\!\!-~}

\usepackage[colorlinks=true,citecolor=blue]{hyperref}

\usepackage{hyperref,cleveref}
\usepackage{color}
\usepackage[dvipsnames]{xcolor}
\usepackage{xcolor}

\begin{document}
\date{\today}
\title{Fermi surface segmentation in the helical state of a Rashba superconductor}

\author{Alireza Akbari$^{1,2}$ and Peter Thalmeier$^1$}
\affiliation{$^1$Max Planck Institute for the  Chemical Physics of Solids, D-01187 Dresden, Germany\\
$^2$Max Planck POSTECH Center for Complex Phase Materials, and Department of Physics, POSTECH, Pohang, Gyeongbuk 790-784, Korea}

\begin{abstract}
We investigate the quasiparticle excitations  in the FFLO- type helical state of a superconductor with inversion-symmetry
breaking and strong Rashba spin-orbit coupling. We restrict to a state with single finite momentum of Cooper pairs in the
helical phase that is determined by minimization of the condensation energy.
We derive the dependence of quasiparticle dispersions  on the Rashba coupling strength and external field. It leads to a peculiar momentum-space segmentation of the corresponding Rashba Fermi surface sheets which has not yet been observed
experimentally. We show that it may be directly visualized by the method of quasiparticle interference that identifies
the critical points of the segmented  sheets and can map their evolution with field strength, bias voltage and Rashba coupling.  We also
indicate a strategy how to determine the finite Cooper-pair momentum from experimental quantities. This investigation has the potential for a more detailed microscopic understanding of the helical superconducting state under the influence of Rashba spin-orbit coupling.
\end{abstract}


\maketitle

\section{Introduction}
\label{sec:introduction}

In a superconductor (SC) with small orbital pair breaking a new state may become stable at larger fields where
the conduction electrons are not bound in BCS pairs $(-\bk\ua,\bk\da)$ but rather in pairs $(-\bk+\bq\ua,\bk+\bq\da)$  with finite center-of-mass momentum $2\bq$ characterized by a gap function $\Delta(\br)=\Delta_\bq\exp(i\bq\br)$. This Fulde-Ferrell-Larkin-Ovchinnikov (FFLO) state~\cite{fulde:64,larkin:65} is well studied theoretically, both in  superconductors of various dimensionality~\cite{shimahara:94,shimahara:98} as well as in condensed quantum gases~\cite{sheehy:07,sheehy:15}.
Convincing evidence for the experimental realization of this state at low temperatures and high fields is, however, rather scarce which may be due to the sensitivity to impurities~\cite{takada:70,matsuda:07,wang:07} and orbital pair breaking~\cite{gruenberg:66,adachi:03}. There are unconventional heavy fermion superconductors~\cite{matsuda:07} and organic superconductors~\cite{lortz:07,mayaffre:14} as well as Fe-pnictides~\cite{burger:13,zocco:13}
where its appearance has been suggested. The existence of the FFLO phase in these cases is mostly infered from thermodynamic anomalies~\cite{bianchi:03} in the low-temperature high-field sector of the phase diagram or from NMR experiments~\cite{kumagai:11} and they may be used to map out the FFLO phase boundaries.\\

However, such experiments do not address the microscopic nature of this state deep inside the FFLO-type phase. The latter is stabilized by a tradeoff between the loss of condensation energy due to the kinetic energy of pairs with center of mass (CM) momentum and gain in Zeeman energy due to population imbalance of spin states~\cite{combescot:07,zwicknagl:11}. This tradeoff depends on the momentum position on the Fermi surface  (FS)  and therefore generally the latter is segmented into regions where the pairs are still stable with finite $2\bq$~(paired region) and where they are unstable (unpaired region). 
The relative size of these FS segments depends on the field strength with the paired region vanishing above the critical field of FFLO phase. This microscopic structure of the FFLO state has not been probed in practice due to lack of suitable techniques. It was proposed in Ref.~\cite{akbari:16}
that STM-based quasiparticle interference (QPI) method is a promising candidate for this purpose. However as a feasibility study only the inversion symmetric superconductor was investigated in this work.\\

In reality inversion symmetry at the superconductor surface is broken and some of the promising SC materials have layered structure with broken 2D inversion symmetry in the layers or even have bulk non-centrosymmetric structure with  complete lack of inversion symmetry. Then Rashba-type spin orbit coupling exists and will greatly modify both the FFLO- type states as well as QPI spectral features. In particular the Fermi surface will be doubled into two Rashba Fermi surfaces with different spin texture.  This important case is therefore worthy of a separate theoretical investigation presented in this work. There is an important distinction, however, to the common FFLO case where the Zeeman term leads to different Fermi sphere radii of up and down spin electrons, whereas under the presence of a dominating Rashba coupling the two Rashba band $(\la=\pm1)$  Fermi spheres are shifted perpendicular to the field by a certain amount proportional to the field strength.  This leads immediately  to stable Cooper pairs with finite momentum $2\bq$ that grows with field strength characterized by an isotropic gap function $\Delta_{\bq\la}\exp(i\bq\br)$.
This commonly called `helical' state~\cite{kaur:05} is therefore of the FFLO type but has a somewhat different composition of the condensation energy than in the original Zeeman dominated FFLO case.\\

Some aspects of the helical state including Rashba coupling and Zeeman term have been studied before, concerning mostly critical field curves~\cite{agterberg:07,loder:13,nakamura:15,zwicknagl:17}.
Here we focus on the microscopic consequences of the Rashba coupling and its image in the QPI spectrum~\cite{Akbari:2013tt}. As a prerequesite we derive the quasiparticle excitation spectrum in the paired and unpaired segments of momentum space whose size depend on the field strength. The corresponding QPI spectrum is created by scattering of quasiparticles from randomly distributed dilute impurities at the surface. We consider normal charge as well as Ising-type magnetic impurities. Our momentum-resolved QPI analysis has a twofold aim: Both the Rashba-doubling of Fermi surface sheets as well as their segmentation  in the helical state due to the appearance of unpaired states may be investigated  as function of field strength, bias voltage, Rashba coupling and chemical potential. In this way one may get a more microscopic understanding of the peculiar helical superconducting state. In particular we show that it is possible to obtain a direct experimental measure of the pair momentum $2\bq$  by analyzing the characteristic wave vectors of the QPI image. The Rashba case with its helical phase is more amenable to such QPI analysis because the latter may appear already at small fields whereas the conventional FFLO phase requires generally very high fields.\\

The model for the Rashba superconductor is introduced in Sec.~\ref{sec:model} and the Bogoliubov quasiparticle excitations are
derived in Sec.~\ref{sec:bogol} following a method introduced by Cui et al in Ref.~\cite{cui:06}  for the inversion-symmetric case without Rashba term (see also Ref.~\onlinecite{akbari:16}). In Sec.~\ref{sec:Green} we calculate the corresponding Green's functions and quasiparticle DOS for the helical phase. In the main part of Sec.~\ref{sec:QPI} we derive the QPI spectrum in Born approximation using a model for impurity scattering that contains both normal and magnetic scattering, transformed to the Rashba band states. Finally Sec. ~\ref{sec:discussion} we discuss the numerical results in detail and Sec.~\ref{sec:conclusion} presents the summary and conclusion.

\section{Model definition}
\label{sec:model}
Here we introduce the commonly used bandstructure model  including the Rashba coupling originating from inversion-symmetry breaking. We use the periodic form in view of the later QPI calculations but sometimes discuss the features of Rashba bands in the 
parabolic approximation for convenience. Subsequently a minimal model for the superconducting s-wave state introduced in Ref.~\cite{kaur:05} will be briefly described and the Hamiltonian for the helical phase discussed.

\subsection{Normal state Rashba bands and states}
\label{subsect:normalmodel}

\begin{figure}
\includegraphics[width=0.99\linewidth]{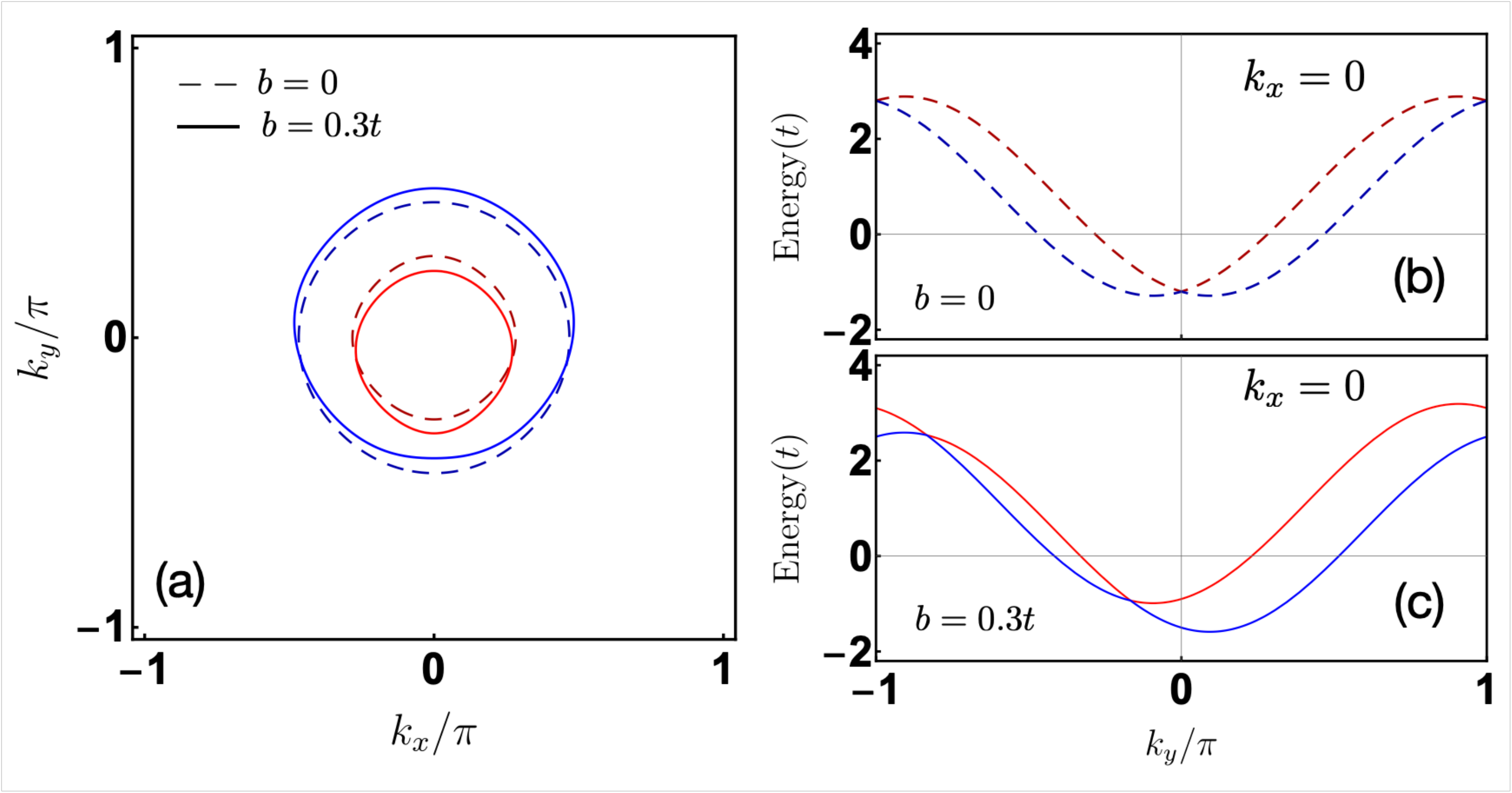}
\caption{
(a) Splitting of Rashba Fermi surface sheets $\sim |\alpha|/\mu$ (dashed) and opposite shifting of their centers by $\pm q_s$ along $k_y$-axis for large field (full); (blue/red for $\lambda=\pm 1$). Magnetic field $\bb\parallel x$- axis and perpendicular shift vector $\bq_s\parallel y$- axis. Here and in the following figures, we set $\mu=-2.8t$ and $\alpha=0.6t$. (b,c) Corresponding TB  dispersions along $k_x$.
}
\label{fig:spectral1}
\end{figure}

The 2D Rashba Hamiltonian in an external field is given by~\cite{kaur:05}
\be
\bl
&
H_{0}=
\sum_\bk\Psi^\dag_\bk h_{0\bk}\Psi_\bk; \;\;\;
 h_{0\bk}=\xi_\bk\sigma_0+(\al\bg_\bk+\bb)\cdot\bsig 
 ,
\label{eq:HRashba}
\el
\ee
in the spin representation. Here $\Psi^\dag_\bk=(a^\dag_{\bk\ua},a^\dag_{\bk\da})$ are conduction electron spinors and 
 $\vare_\bk=-2t(\cos k_x+\cos k_y)$ with $-\pi\leq k_x,k_y\leq \pi$  is the periodic tight binding (TB) conduction band dispersion  which is more suitable for the  later treatment of QPI spectrum. Here $t>0$ is the hopping element leading to a conduction band half-width $D_c=4t$ and $\xi_\bk=\vare_\bk-\mu_{\rm TB}$ . The chemical potential $\mu_{\rm TB}$ in the periodic band model therefore lies in the interval $-D_c\leq \mu_{\rm TB}\leq D_c$ and  is referenced to the {\it band center} $\vare_\bk=0$. It is necessary to map this to the 2D parabolic band model for $\mu_{\rm TB}\leq 0$ with $\vare_\bk=\vare_0+\bk^2/2m$.
  Here $\vare_0=-D_c$ is the bottom of the band and $m=2/D_c$ its effective mass. The chemical potential referenced to the {\it band bottom} is then given by $\mu=\mu_{\rm TB}-\vare_0 \geq 0$.
   Furthermore $\bb=\mu_B{\bf B}$ is the Zeeman energy scale given by the applied magnetic field \bB. The inversion symmetry breaking Rashba spin-orbit coupling is odd under inversion with  $\bg_{-\bk}= -\bg_{\bk}$, explicitly $\bg^P_\bk=(k_y,-k_x,0)/k_F = (\sin\theta_\bk,-\cos\theta_\bk,0)$ in the parabolic band model where $\theta_\bk$ is the azimuthal angle of $\bk$  counted from the $k_x$- axis where the Fermi wave number is  $k_F=(2m\mu)^\fs$ and $v_F=k_F/m$ the Fermi velocity. To stay consistent with the tight binding model dispersion we will take the periodic form
\bea
\bg^{\rm TB}_\bk=(\sin k_y,-\sin k_x,0)
,
\eea
where both forms are normalized, i.e. $|g^P_\bk|=1$ and  $|g^{\rm TB}_\bk|_{\rm max}=\sqrt{2}$. Equivalence in the limit of small wave vectors $k_x,k_y\ll\pi$ demands that the Rashba coupling constants in the two models are then related by $\alpha_P=k_F\alpha_{\rm TB}$.
We suppress indices TB, P in the following and rely on the context.  Diagonalization of the Hamiltonian in Eq.~(\ref{eq:HRashba}) leads to 
\be
\bl
&
H_{0}
=\sum_{\bk\la}\vare_{\bk\la} c^\dg_{\bk\la}c_{\bk\la};\;\;\;
\\
&
\vare_{\bk\la}(\bb)
=\xi_{\bk}+\la|\al\bg_{\bk}+\bb|
\equiv\xi_\bk+\lambda\zeta_\bk^+
,
\label{eq:Rdispb1}
\el
\ee
where $\vare_{\bk\la}(\bb)$ denotes the Rashba- split and Zeeman- shifted bands (refered to $\mu$) which have eigenfunctions corresponding to helicities $\la=\pm 1$. Here we introduce the auxiliary functions $\zeta_\bk^\pm=|\alpha\bg_\bk\pm\bb|$.
In zero field the two Rashba bands are given by
\be
\vare^0_{\bk\la}=\xi_{\bk}+\la |\al\bg_{\bk}| = \frac{1}{2m}(k+\la k_0)^2-\tilde{\mu},
\label{eq:Rdisp0}
\ee
where $k_0=\fs\frac{|\alpha|}{\mu}k_F$ and $\tilde{\mu}=\mu(1+\frac{1}{4}\frac{\al^2}{\mu^2})$. This describes two parabolic dispersions shifted by $k_0$. The ensuing two Fermi spheres have radii given approximately by $k_F^\la=k_F-\la k_0=k_F(1-\frac{\la}{2}\frac{|\alpha|}{\mu})$ for  moderate Rashba coupling $|\alpha|\ll\mu$. Then their relative difference $(k_F^--k_F^+)/k_F=|\alpha|/\mu$ is a direct measure for the strength of the Rashba coupling.
The operators $\Phi_\bk^\dag=(c_{\bk +}^\dag c^\dag_{\bk -})$ $(\la=\pm)$ creating the helical eigenstates $|\bk\la\rangle=c^\dag_{\bk\la}|0\rangle$ are obtained~\cite{thalmeier:20} from 
\bea
\Phi^\dag_\bk=\Psi^\dag_\bk S_\bk; \;\;\;
S_\bk=
\frac{1}{\sqrt{2}}
\left[
\begin{matrix}
1& ie^{-i\theta_\bk} \cr
 ie^{i\theta_\bk} &  1
 \end{matrix}
 \right],
 \label{eq:helical}
\eea
where $\theta_\bk=-\tan^{-1}(g_{\bk x}/g_{\bk y})=\tan^{-1}(\sin k_y/\sin k_x)\rightarrow \tan^{-1}(k_y/k_x)$ where the second and last expression correspond to TB and parabolic models, respectively. For finite but small in-plane field $\bb=b(\cos\phi,\sin\phi,0)$
 $(b\ll\alpha)$, where $\phi$ defines the field angle with respect to planar axes
the Rashba dispersions of Eq.~(\ref{eq:Rdispb1}) in the parabolic model can be written  explicitly as
\be
\bl
\vare_{\bk\la}(\bb) = 
&
\frac{1}{2m}(k+\la k_0)^2-\tilde{\mu}
+\la s_\al b\sin(\theta_\bk-\phi),
\el
\ee
where we defined $s_\alpha={\rm sign}(\alpha)$. This leads to  Rashba Fermi sheets with a radius
given by 
\bea
k_F^\la(\theta_\bk,\phi)=k^\lambda_F
-
\frac{1}{2\mu} k_F\la s_\alpha b \sin(\theta_\bk-\phi),
\eea
%
%
where
 we assumed the physical hierarchy of energy scales according to  $(b < |\alpha| < \mu < 2D_c)$. An example for the geometry
of Rashba Fermi surface sheets is shown in Fig.~\ref{fig:spectral1}. The effect of the field on the two Rashba sheets may be easily
understood by considering the relative change compared to the zero-field value $k^\lambda_F$ as function of the angle $\theta_\bk$.
For momentum (anti-)parallel to the field with $\theta_\bk=\phi+\pi,\phi$ there is no change and $k_F^\la(\theta_\bk,\phi)=k^\lambda_F$. For perpendicular case with
$\theta_\bk=\phi+\pi/2,\phi+3\pi/2$, 
we have 
$k_F^\la(\theta_\bk,\phi)=k^\lambda_F \mp \fs k_F\lambda s_\alpha(b/\mu)$.
Thus the two Rashba sheets are shifted {\it perpendicular} to the field in opposite directions by the amount
\bea
q_s=
\frac{b}{2\mu}k_F=\frac{m\mu_BB}{k_F}=\frac{b}{v_F}.
\label{eq:Rshift}
\eea
While the splitting of Rashba sheets is a measure for the coupling $|\alpha|$ their shifting perpendicular to \bB~is a determined by field
strength alone. These basic Rashba characteristics are shown in Fig.~\ref{fig:spectral1}(a) as it results from the splitted and shifted dispersions in  Figs.~\ref{fig:spectral1}(b,c).

\subsection{Superconducting state with finite momentum pairing}
\label{subsect:supermodel}

In this work we do not discuss the possible mechanisms behind the superconducting gap formation in non-centrosymmetric compounds without inversion symmetry, for an excellent review see Ref.~\cite{sigrist:09}. In these materials with Rashba spin-oribit coupling phonons~\cite{wiendlocha:16} as well as spin-fluctuations~\cite{yanase:08,takimoto:09,mukherjee:12} may be the driving mechanism for Cooper pair formation. In any case it is important to realize
that, independent of the mechanism the gap function contains spin- singlet as well as triplet components due to
the inversion symmetry breaking presented by the Rashba term.

%
%

In addition here we consider the possibility of a common overall momentum $2\bq$ of Cooper pairs due
to the pairbreaking effect of the external field in conjunction with Rashba spin-orbit coupling. One should expect
that the size of $\bq$ is correlated with the shift of the Rashba FS sheets~\cite{agterberg:07} perpendicular to the field as given
by Eq.~(\ref{eq:Rshift}). The real value of \bq~should be evaluated by the minimization of the condensation
energy in the helical SC phase as is demonstrated in Sec.~\ref{sec:bogol}.
As mentioned in the introduction more general pairs with multiple $\bq_i$, in particular the 
`stripe phase'~\cite{kaur:05} with $(\bq,-\bq)$ will not be considered here. Of the many possible choices of gap functions we use the minimal model introduced by Kaur et al~\cite{kaur:05} which reduces to 
the spin-singlet form in the limit $\alpha=0$. In the helical basis it is characterized by two gap functions $\Delta^\bk_{\bq\la}$ for the two Rashba sheets. The resulting mean field pair  Hamiltonian in helicity representation is described by
\be
\bl
H_{\rm MF}
=&
\sum_{\bk\la}\vare_{\bk+\bq\la} c^\dg_{\bk+\bq\la}c_{\bk+\bq\la}
\\
&-
\frac{1}{2}\sum_{\bk\la}
[
\Delta_{\bq\la}c^\dg_{\bk+\bq\la}c^\dg_{-\bk+\bq\la}
+\Delta^*_{\bq\la}c_{-\bk+\bq\la}c_{\bk+\bq\la}
].
\label{eq:hmf}
\el
\ee
With the gap equation for the isotropic state of the FFLO phase given by
\bea
\Delta_{\bq\la}=-\sum_{\bk '\la'}{'}V_{\la\la'}\langle c_{-\bk'+\bq\la '}c_{\bk '+\bq\la '}\rangle
,
\label{eq:gapeq0}
\eea
where the prime indicates that the summation over $\bk',\la'$ runs only over the paired momentum regions with {\it positve} quasiparticle energies in Eq.~(\ref{eq:SCquasi}). For the singlet case (in the limit $\alpha=0$)  considered the pairing interaction in the helical basis in the limit  $|\bb/\alpha | \ll 1$ takes the form~\cite{kaur:05} 
\bea
\hV=-  \frac{V_0}{2}(\sigma_0-\sigma_x)=
-\frac{V_0}{2}
\left[
 \begin{array}{cc}
 1& -1 \\
 -1& 1
\end{array}
\right].
\label{eq:vmat0}
\eea
Inserting this two-band pairing interaction into the gap equation [Eq.~(\ref{eq:gapeq0})] leads to the condition $\Delta^\bk_{\bq-}=-
\Delta^\bk_{\bq+}$. The opposite sign of the two gaps is enforced by the opposite spin texture on the two Rashba bands.
To keep the parameter set for our investigation at a minimum we restrict to the simple case of {\it isotropic} ($\Delta_{\bq-}=-\Delta_{\bq+}$) gaps without \bk - dependence. Then the total BCS Hamiltonian including the mean field energy constant is given by
\bea
H_{\rm BCS}=H_{\rm MF}+\frac{1}{2}\sum_{\bk\la}\frac{\Delta^2_{\bq\la}}{V_0}.
\label{eq:hbcs}
\eea
Introducing the Nambu spinors $\psi^\dg_{\bk\bq\la}=(c^\dg_{\bk+\bq\la},c_{-\bk+\bq\la})$ we may write $H_{\rm MF}=\hat{H}_{\rm MF}+E_0$ where
\be
\bl
\hat{H}_{\rm MF}
=\frac{1}{2}\sum_{\bk\la}\psi^\dg_{\bk\bq\la}\hh_{\bk\bq\la}\psi_{\bk\bq\la}; \;\;\;
E_0=\frac{1}{2}\sum_{\bk\la}\vare_{\bk+\bq\la},
\el
\ee
and  the Hamilton matrix in Nambu space is given by
\be
\bl
\hh_{\bk\bq\la}=
\left[
 \begin{array}{cc}
 \vare_{\bk+\bq\la}&-\Delta_{\bq\la} \\
 -\Delta_{\bq\la}^*& -\vare_{-\bk+\bq\la}
\end{array}
\right].
\label{eq:hmat}
\el
\ee
Using the symmetries $\xi_\bk=\xi_{-\bk}$ and $\bg_\bk=-\bg_{-\bk}$ the diagonal elements are obtained as
\be
\bl
\vare_{\bk +\bq\la}(\bb)
&=
\xi_{\bk +\bq}+\la|\al\bg_{\bk+\bq}+\bb|
=\xi_{\bk +\bq}+\la\ze^+_{\bk\bq};
\\
\vare_{-\bk +\bq\la}(\bb)
&=
\xi_{\bk -\bq}+\la|\al\bg_{\bk-\bq}-\bb|
=\xi_{\bk -\bq}+\la\ze^-_{\bk\bq},
\el
\ee
here we defined the auxiliary Rashba functions
$\ze^\pm_{\bk\bq} = |\al\bg_{\bk\pm\bq}\pm\bb|$ where both signs on the right are taken simultaneously $+$ or $-$.
We also introduce symmetric (s) and antisymmetric (a) combinations explicitly given by
%
%
%
\be
\bl
\vare^{s}_{\bk\bq\la}
&=
\frac{1}{2}(\vare_{\bk +\bq\la}+\vare_{-\bk +\bq\la})
\\&=
\frac{1}{2}(\xi_{\bk+\bq}+\xi_{\bk-\bq})+
\la\frac{1}{2}(\ze^+_{\bk\bq}+\ze^-_{\bk\bq})
\equiv \xi^s_{\bk\bq}+\la\ze^s_{\bk\bq},
\\
\vare^{a}_{\bk\bq\la}
&=\frac{1}{2}(\vare_{\bk +\bq\la}-\vare_{-\bk +\bq\la})
\\&=
\frac{1}{2}(\xi_{\bk+\bq}-\xi_{\bk-\bq})+
\la\frac{1}{2}(\ze^+_{\bk\bq}-\ze^-_{\bk\bq})
\equiv \xi^a_{\bk\bq}+\la\ze^a_{\bk\bq},
\label{eq:symeps}
\el
\ee
They fulfil the even/odd symmetry relations $\vare^{s}_{-\bk\bq\la}=\vare^{s}_{\bk\bq\la}$ and 
$\vare^{a}_{-\bk\bq\la}=-\vare^{a}_{\bk\bq\la}$, respectively. Here we defined  $\xi^{s,a}_{\bk\bq}=
\frac{1}{2}(\xi_{\bk+\bq}\pm\xi_{\bk-\bq})$ and  $\ze^{s,a}_{\bk\bq}=\frac{1}{2}(\ze^+_{\bk\bq}\pm\ze^-_{\bk\bq})$.
In the formal limit of no Rashba coupling ($\alpha=0$) this simplifies to  $\ze^{s}_{\bk\bq}=|b|$ and  $\ze^{a}_{\bk\bq}=0$.
In this case the two Rashba bands $\epsilon_{\bk\la}$ (Eq.~(\ref{eq:Rdispb1}))  become the Zeeman split bands with effective spin
index $\la$.
Now we can split the diagonal matrix elements in the Hamiltionian into symmetric and antisymmetric parts and using
the symmetry relations we arrive at
\be
\bl
\hh_{\bk\bq\la}=
\vare^a_{\bk\bq\la}\tau_0+
\left[
 \begin{array}{cc}
 \vare^s_{\bk\bq\la}&-\Delta_{\bq\la} \\
 -\Delta_{\bq\la}^*& -\vare^s_{\bk\bq\la}
\end{array}
\right]
.
\label{eq:hsmat}
\el
\ee
We note that in the following we
will also use the property $\sum_{\bk\la}\vare^a_{\bk\bq\la}=0$ which is due the antisymmetry of $\vare^a_{\bk\bq\la}$. Hereby the summation over \bk~ runs over the paired and unpaired regions as defined below.

\section{Bogoliubov transformation for paired and depaired states}
\label{sec:bogol}

The first part in the  \bk - symmetrized Hamiltionian in Eq.~(\ref{eq:hsmat}) is already diagonal. 
The second part can now be diagonalized by a Bogoliubov transformation to quasiparticle states created
by $\alpha_{\bk\la},\beta_{\bk\la}$ with the corresponding Hamiltonian expressed as
\be
\bl
H_{\rm MF}=
&
\frac{1}{2}\sum_{\bk\la}\bigl[|E^+_{\bk\bq\la}|\al^\dg_\bk\al_\bk+|E^-_{\bk\bq\la}|\bt^\dg_\bk\bt_\bk\bigr]
\\
&+
\frac{1}{2}\sum_{\bk\la}{'}
\left\{
 \begin{array}{c}
 \vare^s_{\bk\bq\la}-E_{\bk\bq\la};\;\;E^\tau_{\bk\bq\la}>0\\
\vare^s_{\bk\bq\la}+\vare^a_{\bk\bq\la};\;\;E^+_{\bk\bq\la}<0\\
\vare^s_{\bk\bq\la}-\vare^a_{\bk\bq\la};\;\;E^-_{\bk\bq\la}<0
\end{array}
\right\}.
\quad
\label{eq:hmfbogol}
\el
\ee
Here the quasiparticle energies are given by $(\tau=\pm,\bar{\tau}=\mp)$: 
\bea
\bl
E^\tau_{\bk\bq\la}
&=E_{\bk\bq\la}+\tau\vare^a_{\bk\bq\la}=E^{\bar{\tau}}_{-\bk\bq\la},
\\
E_{\bk\bq\la}
&=[\vare^{s2}_{\bk\bq\la}+\Delta_{\bq\la}^2]^\frac{1}{2}=E_{-\bk\bq\la}.
\label{eq:SCquasi}
\el
\eea
When for a given $\bk, \la$ both $E^\tau_{\bk\bq\la}>0$  $(\tau=\pm)$ one has a stable Cooper pair state with pair momentum $2\bq$ for this wave vector $\bk$ and band $\la$. If, on the other hand $E^+_{\bk\bq\la}<0$ or $E^-_{\bk\bq\la}<0$ the pair state is broken  and only unpaired quasiparticle states at exist at the wave vectors $\bk+\bq$, $-\bk+\bq$. Note the remarkable fact that although for these wave vectors $|E^\pm_{\bk\bq\la}|$ are normal quasiparticle excitations their energy nevertheless contains the gap size $\Delta_{\bq\la}$ determined by the paired states. This is because in the coherent helical ground state the unpaired electrons and holes also experience the pairing molecular field sustained by the paired electrons, even though they do not contribute to it.
The mean field energy constant (last term in Eq.~(\ref{eq:hmfbogol})) in the two cases is different because of the additional condensation energy in the paired state.\\

Therefore  the corresponding Bogoliubov transformations for the two cases are also different:
For the paired states it is given by ~\cite{cui:06}:
\be
E^\tau_{\bk\bq\la}>0:
\left[
 \begin{array}{c}
 c_{\bk+\bq\la}\\
 c^\dg_{-\bk+\bq\la}
\end{array}
\right]
=
\left[
 \begin{array}{cc}
 u^*_{\bk\la}&v_{\bk\la}\\
 -v^*_{\bk\la}&u_{\bk\la}
\end{array}
\right]
\left[
 \begin{array}{c}
 \alpha_{\bk\la}\\
 \beta^\dg_{\bk\la}
\end{array}
\right],
\label{eq:BGLp}
\ee
whereas for the depaired states it may be written as~\cite{cui:06}
\bea
\bl \non
&
E^+_{\bk\bq\la}<0:
\left[
 \begin{array}{c}
 c_{\bk+\bq\la}\\
 c^\dg_{-\bk+\bq\la}
\end{array}
\right]
=
\left[
 \begin{array}{cc}
 u^*_{\bk\la}&v_{\bk\la}\\
 -v^*_{\bk\la}&u_{\bk\la}
\end{array}
\right]
\left[
 \begin{array}{c}
 \alpha^\dg_{\bk\la}\\
 \beta^\dg_{\bk\la}
\end{array}
\right]
;\;\;\;
\\&
E^-_{\bk\bq\la}<0:
\left[
 \begin{array}{c}
 c_{\bk+\bq\la}\\
 c^\dg_{-\bk+\bq\la}
\end{array}
\right]
=
\left[
 \begin{array}{cc}
 u^*_{\bk\la}&v_{\bk\la}\\
 -v^*_{\bk\la}&u_{\bk\la}
\end{array}
\right]
\left[
 \begin{array}{c}
 \alpha_{\bk\la}\\
 \beta_{\bk\la}
\end{array}
\right].
\label{eq:BGLu}
\el
\\
\eea

Explicitly the transformation coefficients are given by
\bea
 u_{\bk\la}^2&=&\frac{1}{2}\bigl(1+\frac{\vare^s_{\bk\bq\la}}{E_{\bk\bq\la}}\bigr); \;\;\;
 v_{\bk\la}^2=\frac{1}{2}\bigl(1-\frac{\vare^s_{\bk\bq\la}}{E_{\bk\bq\la}}\bigr).
\eea
These coefficients fulfil the wellknown relations
\bea
u_{\bk\la}^2-v^2_{\bk\la}=\frac{\vare^s_{\bk\bq\la}}{E_{\bk\bq\la}};\;\;
2u_{\bk\la} v_{\bk\la}=\frac{\Delta_{\bq\la}}{E_{\bk\bq\la}}.
\eea
Note the important fact that only the { \it symmetrized} Rashba band energies $\vare^s_{\bk\bq\la}$ appear in the transformation coefficients $u_{\bk\la}, v_{\bk\la}$. However, both 
momentum- symmetric $\vare^s_{\bk\bq\la}$ and  anti- symmetric  $\vare^a_{\bk\bq\la}$ contribute to the superconducting quasiparticle energies $E^\tau_{\bk\bq\la}$ in Eq.~(\ref{eq:SCquasi}). This result of the analysis could not have been anticipated a priori with heuristic arguments.\\

\begin{figure}
\includegraphics[width=0.99\linewidth]{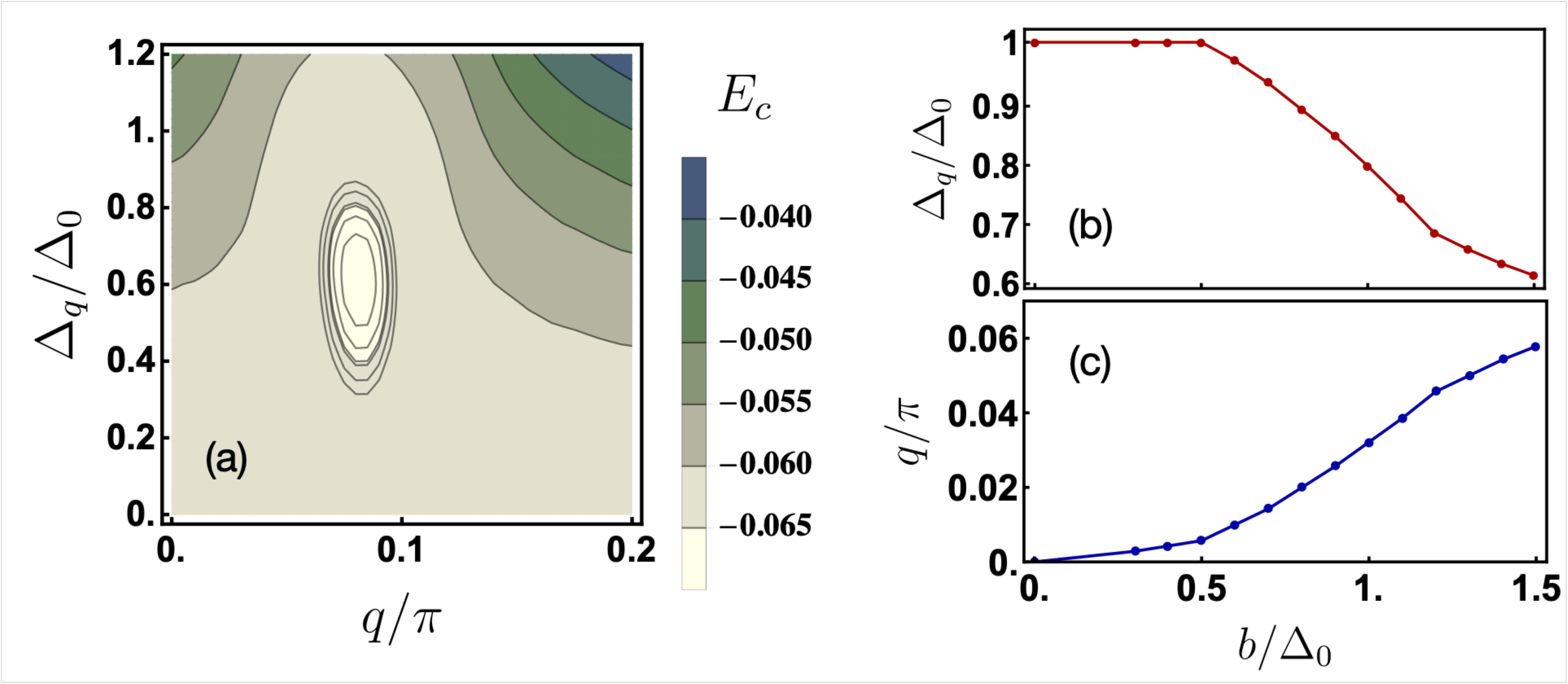}
\caption{
{
(a) Contour plots of SC condensation energy $E_c <0$ in the helical state in the $(q,\Delta_q)$- plane
for typical field $\bb=\mbox{b}\hbx$ with $b<\alpha$.  
(b,c) Field dependence of helical gap size $\Delta_\bq$, and  (half-) pair momentum $\bq=\mbox{q}\hby$, 
 corresponding to minimum in (a).
Here and  in the rest of the paper, we set $\Delta_0=0.3t$, $\alpha=2\Delta_0=0.6t$.
}
}
\label{fig:Etot}
\end{figure}

The total BCS Hamiltionian, including the constant energy in Eq.~(\ref{eq:hbcs}) is then obtained as 
\bea
\bl
H_{\rm BCS}
= &
H_{\rm MF}
+\frac{1}{2}\sum_{\bk\la}{'}\frac{|\Delta|^2_{\bq\la}}{V_0}
\non
\\
= &
\frac{1}{2}\sum_{\bk\la}(|E^+_{\bk\bq\la}|\al^\dg_\bk\al_\bk+|E^-_{\bk\bq\la}|\bt^\dg_\bk\bt_\bk)
\\
&+
\frac{1}{2}\sum_{\bk\la}{'}
\left\{
 \begin{array}{c}
 \vare^s_{\bk\bq\la}-E_{\bk\bq\la}+\frac{|\Delta|^2_{\bq\la}}{V_0}\\
\vare^s_{\bk\bq\la}+\vare^a_{\bk\bq\la}+\frac{|\Delta|^2_{\bq\la}}{V_0}\\
\vare^s_{\bk\bq\la}-\vare^a_{\bk\bq\la}+\frac{|\Delta|^2_{\bq\la}}{V_0}
\end{array}
\right\}
.
\label{eq:bcsbogol}
\el
\\
\eea
Here the second term $\langle H_{\rm BCS}\rangle$  is equal the total ground state  energy $E_G(\bq,\Delta_{\bq\pm})$ of the helical FFLO-type state. As in Eq.~(\ref{eq:hmfbogol}) the sum extends over the upper value for paired states with both $E^\pm_{\bk\bq\la}>0$ whereas the lower
values correspond a sum only over to the unpaired states with  $E^+_{\bk\bq\la}<0$ or  $E^-_{\bk\bq\la}<0$, respectively. These conditional sums are indicated by the prime. The helical SC ground state energy may be rewritten explicitly as (see also Appendix \ref{sec:GSenergy})
\bea\non
E_G(\bq,\Delta_{\bq\pm})
&=&
\fs\sum_\la\Bigl[
N\bigl(\frac{|\Delta_{\bq\la}|^2}{V_0}\bigr)
-\sum_\bk(E_{\bk\bq\la}-\vare^s_{\bk\bq\la})
\\&&\hspace{-2cm}
+\sum_\bk[E^+_{\bk\bq\la}\Theta(-E^+_{\bk\bq\la})+E^-_{\bk\bq\la}\Theta(-E^-_{\bk\bq\la})]\Bigr].
\label{eq:GSen}
\eea
This energy functional should be minimized with respect to \bq~and $\Delta_{\bq\pm}$ for Rashba coupling $\alpha$ and field strength $b$.
It contains the possibilities of the helical $(\bq\neq 0,|\Delta_{\bq\la}|>0)$, BCS $(\bq=0,|\Delta_{0\la}|>0)$ and unpolarized normal $(b=0,\bq=0, \Delta_{\bq\la}=0)$ states. For the latter the ground state energy is 
\bea
\bl
E^0_G
=&
\fs\sum_{\bk\la}(\vare^0_{\bk\la}-|\vare^0_{\bk\la}|)=\sum_{\bk\la}f_{\bk\la}\vare^0_{\bk\la},
\\
\vare^0_{\bk\la}
=&\vare^s_{\bk\bq\la}(\bq=0,b=0)=\xi_\bk+\la|\alpha\bg_\bk|,
\label{eq:gsen}
\el
\eea
where $f_\bk=\Theta(-\vare^0_{\bk\la})$ is the zero temperature Fermi function for the unpolarized Rashba-split bands $\vare^0_{\bk\la}$ (cf. Eq.~(\ref{eq:Rdisp0})). The minimization problem is greatly simplified  by the equal size of the gaps $|\Delta_{\bq\pm}|=\Delta_\bq$ in the model defined by Eq.~(\ref{eq:vmat0}). Although strictly this holds only for \bq=0 we will also keep this minimization constraint for the helical case. The pairing potential strength $V_0$ in Eqs.~(\ref{eq:vmat0},\ref{eq:gsen}) is related to the gap size $\Delta_0$ by the simplified single gap equation obtained from  Eq.~(\ref{eq:gapeq0})
\bea
\frac{1}{V_0}=\frac{1}{2N}\sum_{\bk\la}\frac{1}{2E_{\bk\la}}\Theta(\vare_c-|\vare_{\bk\la}|),
\label{eq:pairpot}
\eea
where the BCS zero-field quasiparticle energy is $E_{\bk\la}=[\vare^{02}_{\bk\la}+\Delta_0^2]^\fs $. Here $\xi_c$ is an effective cutoff
of the pairing potential $(\Delta_0<\xi_c<2D_c)$. In the following calculations the gap size $\Delta_0$ is used directly as a fixed input parameter, then the cutoff may be absorbed in an effective coupling constant $V_0$ by deleting the $\Theta$- function.\\
For finding the ground state by numerical minimization it is useful to subtract the normal state energy from the ground state energy in Eq.~(\ref{eq:GSen}) to obtain the superconducting condensation energy $E_c=E_G-E^0_G$ according to
\bea
\bl
&
E_c(\bq,\Delta_{\bq\pm})=
\fs\sum_\la
\Bigg[
N\bigl(\frac{|\Delta_{\bq\la}|^2}{V_0}\bigr)
-\sum_\bk
\Big[(E_{\bk\bq\la}-|\vare^0_{\bk\la}|)
\non\\
&
\!
+(\vare^s_{\bk\bq\la}
\!-\!
\vare^0_{\bk\la})
+[E^+_{\bk\bq\la}\Theta(-E^+_{\bk\bq\la})+E^-_{\bk\bq\la}\Theta(-E^-_{\bk\bq\la})]
\Big]
\Bigg].
\el
\\
\label{eq:condens}
\eea
Note that the asymmetric $\vare^a_{\bk\bq\la}$ Rashba energies of Eq.~(\ref{eq:symeps}) enter only in the unpaired quasiparticle contribution (last term).
Using the pure singlet gap constraint  $\Delta_{\bq\pm}=\pm\Delta_\bq$ the minimization of $E_c(\bq,\Delta_{\bq})$  with respect to $\Delta_{\bq}$ and \bq~ for fixed field \bb~ and Rashba coupling $\alpha$ determines the equilibrium gap $\Delta(\bq,b,\al)$ and wave vector  $\bq(b,\al)$ characterizing the helical state. We have to keep in mind, however, that the pairing model of Eq.~(\ref{eq:vmat0}) is only strictly valid in the low field limit $b/\alpha\ll1$.  An example of the condensation energy minimum formation in the $(q,\Delta_q)$ plane and the resulting $\Delta_q(b), q(b)$ dependence for small fields and fixed $\alpha$ is shown in Fig.~\ref{fig:Etot}.

%
%

\section{Green's functions in the Rashba-FFLO state and quasiparticle DOS}
\label{sec:Green}

\begin{figure}
\includegraphics[width=0.950\linewidth]{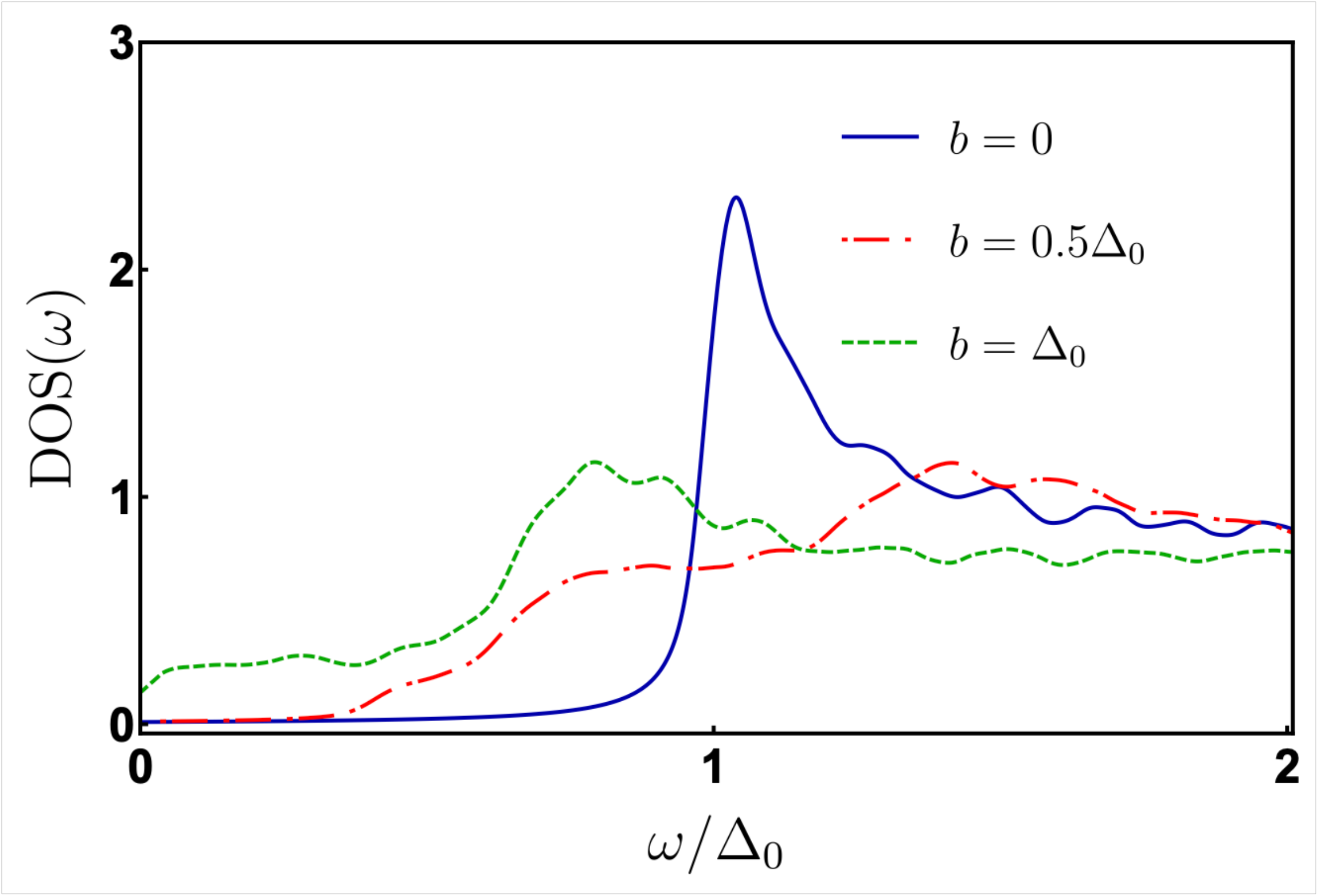}
\caption{
Comparison of quasiparticle DOS, $\rho_{\bq}(\omega)$,  in Rashba-BCS state $(b=0, q=0, \Delta_0)$
and helical state $(b,q\neq 0, \Delta_q)$. The low energy DOS appears due to
normal quasiparticles in the unpaired momentum space region (cf. Fig.~\ref{fig:spectral2}).
 }
\label{fig:DOS}
\end{figure}

\begin{figure}
\includegraphics[width=\linewidth]{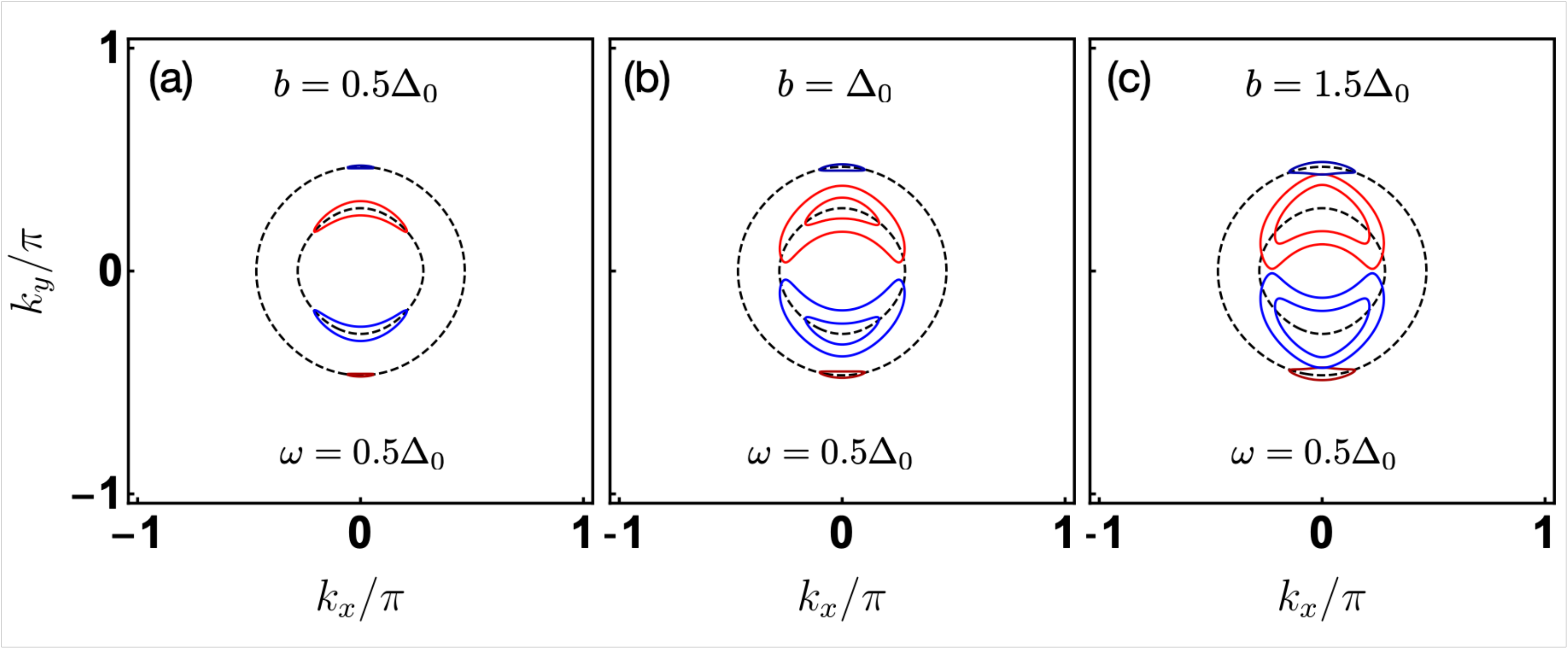}
\caption{
Evolution of spectral function in Eq.~(\ref{eq:specfunc}) with field b at frequency $\omega=0.5\Delta_0$:
a) $b=0.5\Delta_0$ with $q(b)/\pi=0.005 $, and $\Delta_q(b) = \Delta_0$;
b) $b=\Delta_0$ with $q(b)/\pi=0.035 $, and $\Delta_q (b)= 0.75\Delta_0$;
c) $b=1.5\Delta_0$ with $q(b)/\pi=0.06 $, and $\Delta_q(b) = 0.59\Delta_0$.
The dashed lines are corresponding to the bare (normal state) Rashba contours at $\omega=0$ in zero field  (Eq.~(\ref{eq:Rdisp0})).
}
\label{fig:spectral2}
\end{figure}

The Green's functions in the FFLO- type superconducing state are needed for the calculation of quasiparticle DOS and interference spectra. Using Eq.~(\ref{eq:hsmat})
we obtain:
\be
\bl
&
\hG_{\bq\la}(\bk,\om) =
(\om-\hh_{\bk\bq\la})^{-1}
\\
 &
 \hspace{1cm}=
 \frac{1}{(\om-E^+_{\bq\bk\la})(\om+E^-_{\bq\bk\la})}
 \times
 \\
 &
  \hspace{1.42cm}
\left[
 \begin{array}{cc}
\om+\vare^s_{\bk\bq\la}- \vare^a_{\bk\bq\la}
&-\Delta_{\bq\la} \\[0.2cm]
-\Delta_{\bq\la}^*
 &\om -\vare^s_{\bk\bq\la}- \vare^a_{\bk\bq\la}
\end{array}
\right].
\label{eq:gmat}
\el
\ee
The normal and anomalous  Green's functions elements $G^{\tau,\tau'}_{\bq\la}(\bk,\om)$ satisfy the following symmetry relations:
\bea
G^{11}_{\bq\la}(-\bk,\om)=-G^{22}_{\bq\la}(\bk,-\om), \;\;\;
\eea
and likewise
\bea
G^{12*}_{\bq\la}(-\bk,\om)=G^{21}_{\bq\la}(\bk,-\om). \;\;\;
\eea
The spectral function corresponding to the above Green's function is obtained as
\bea
\bl
A^\la_{\bk\bq}(\omega)
&=-\frac{1}{\pi}
{\rm Im} [{\rm tr}
[\hG_{\bq\la}(\bk,\omega+i\eta)
]
]_{\eta\rightarrow 0^{+}}
\\
&=
\sum_\tau\delta(\omega-E_{\bq\bk\la}^\tau).
\label{eq:greenspec1}
\el
\eea
Now using the symmetry relation $E^\tau_{-\bk\bq\la}=E^{\bar{\tau}}_{\bk\bq\la}$ (with $\tau=\pm$ and $\bar{\tau}=\mp$)
one can define symmetrized spectral function according to
\be
\bl
\bar{A}^\la_{\bk\bq}
=
\fs
[A^\la_{\bk\bq}(\omega)+A^\la_{-\bk\bq}(\omega)
]
=
\fs
[
A^\la_{\bk\bq}(\omega)+A^\la_{\bk\bq}(-\omega)
].
\el
\ee
Using Eq.~(\ref{eq:greenspec1}) they may be obtained for paired as well as unpaired regions as
\be
\bar{A}^\la_{\bk\bq}(\omega)=\fs\sum_\tau[\delta(\omega-|E^\tau_{\bk\bq\la}|)+
\delta(\omega+|E^\tau_{\bk\bq\la}|].
\label{eq:specfunc}
\ee
This result agrees with the expression that may be directly infered from the quasiparticle Hamiltonian of Eq.~(\ref{eq:hmfbogol}).
Summation over quasiparticle momenta \bk~then leads to the quasiparticle DOS, $\rho_{\bq\la}(\omega)$,
for Rashba band $\lambda$ in the helical state with pair momentum $2\bq$ according to
\be
\rho_{\bq\la}(\omega>0)=\frac{1}{2N}\sum_{\bk}[ \delta(\omega -|E^+_{\bk\bq\la}|)+\delta(\omega -|E^-_{\bk\bq\la}|)].
\label{eq:qpDOS}
\ee
This presentation for the DOS is perfectly adequate for its numerical evaluation and will in fact be used later. However, to elucidate 
the distinction between conduction bands without spin-orbit coupling $(\alpha =0)$ treated previously~\cite{cui:06,akbari:16} and the present Rashba-split
bands it is illuminating to evaluate this expression partly analytically, except for a remaining momentum angle integration.
For that purpose we can simplify the expressions in Eq.~(\ref{eq:symeps}) when $q/k_F \ll1$ and  $\bk\simeq k_F\hat{\bk}$ is close to the Fermi surface. Then $\xi_\bk^s	\simeq \xi_\bk$ and $\xi^a_\bk\simeq qv_F\cos(\theta_\bk-\theta_\bq)$ with $v_F=k_F/m$ and the orthogonal pair momentum and field directions defined by $\theta_\bq=\frac{\pi}{2}$ $(\bq=\mbox{q}\hat{{\bf y}})$ and $\bb=\mbox{b}\hat{{\bf x}}$.
Furthermore this leads to $\bg_{\bk+\bq}\simeq \bg_\bk \simeq (k_F^y,-k_F^x,0)/k_F=(\sin\theta_\bk,-\cos\theta_\bk,0)$.
Using these approximations we get $(\xi_\bk=\bk^2/2m-\mu)$
\be
\bl
&
\vare^s_{\bk\bq\la}\equiv\vare_{\bk\la}=\xi_\bk+\la | \al\bg_\bk |=\xi_\bk+ \la | \al | ,
\\
&\vare_{\bk\bq\la}^a=(v_Fq+\la b)\sin\theta_\bk.
\el
\ee
In this approximation the superconducting quasiparticle energies (Eq.~\ref{eq:SCquasi}) then simplify to
\be
E^\tau_{\bk\bq\la}
=[(\xi_{\bk}+\la |\al |)^2+\Delta_{\bq\la}^2]^\frac{1}{2}+\tau(v_Fq+\la b)\sin\theta_\bk.
\label{eq:qpera2}
\ee
The quasiparticle DOS $\rho(\omega)$ may be evaluated~\cite{cui:06,akbari:16} as
\be
\bl
\rho_{\bq}(\omega>0)=
&
\frac{1}{4\pi}\sum_\la\rho^n_\la(0)
\int_0^{2\pi}
\!
d\theta
\int_0^{\hbar\omega_c}
\!
d\vare
\\&
\Big[
\delta(\omega-|E^+_{\bk\bq\la}|)+|\delta(\omega-|E^-_{\bk\bq\la}|)
\Big],
\quad\quad
\label{eq:qpdos2}
\el
\ee
where $\rho^n_\la(0)=\rho_n(0)=m/2\pi$ is the normal state DOS equal for both Rashba bands. With the angle-independent
bare Rashba dispersion $\vare^0_{\bk\la}=\xi_{\bk}+\la |\al |$ denoted by $\vare$ we can write
\be
E^\tau_{\bk\bq\la}=[\vare^2+\Delta_{\bq\la}^2]^\frac{1}{2}+\tau(v_Fq+\la b)\sin\theta_\bk.
\ee
Introducing now $\hE^\tau_{\theta q\la}=\omega-\tau(v_Fq+\la b)\sin\theta$,  the $\vare$ - integration leads to the partial radial DOS at angle $\theta=\theta_\bk$
\be
\hat{\rho}_{\bq\la}(\omega,\theta)=\frac{1}{2}\Bigl(
\frac{|\hE^+_{\theta q\la}|}{[|\hE^+_{\theta q\la}|^2-\Delta^2_{q\la}]^\frac{1}{2}}+
\frac{|\hE^-_{\theta q\la}|}{[|\hE^-_{\theta q\la}|^2-\Delta^2_{q\la}]^\frac{1}{2}}\Bigr),
\ee
and the total DOS is then given by 
\be
\rho_\bq(\omega)=\frac{1}{2\pi}\sum_\la\rho^n_\la(0)
\int_0^{2\pi}d\theta\hat{\rho}_{\bq\la}(\omega,\theta).
\ee
Which has four contributions  due to two quasiparticle branches for each of the two Rashba split bands characterized
by $(\tau,\la)=(\pm,\pm)$. They have the same form and are determined by their different energies $\hE^\tau_{\theta q\la}$ which
are explicitly given by
\be
\bl
\hE^\pm_{\theta q +}= \omega\mp(v_Fq+b)\sin\theta;\;\;
\hE^\pm_{\theta q -}= \omega\mp(v_Fq-b)\sin\theta.
\el
\ee
Note that because of the helical spin polarization of Rashba states the Zeeman contribution for a fixed field direction is now also proportional to $\sin\theta$ since the spins are locked with respect to crystal axes for $|\alpha|\gg |\bb|$. This is an essential difference to the inversion symmetric case without Rashba spin-orbit coupling where they can align parallel to the \bb- field~\cite{cui:06,akbari:16} and therefore no dependence on the momentum angle $\theta$ appears in this case.
An example of the quasiparticle DOS, using the general form of Eq.~(\ref{eq:qpDOS}) is shown in Fig.~\ref{fig:DOS}. As the field increases and unpaired states appear in the helical phase the corresponding low energy normal quasiparticles gradually fill up
the SC gap. It is important to note that a zero energy quasiparticle DOS appears although the helical SC order parameter has no nodes, neither in \bk- space nor in real space. This is rather a consequence of  the presence of FS sheets of unpaired states defined by $|E^\tau_{\bk\bq\la}|=\omega$. Their evolution with field b for constant frequency is  shown in Fig.~\ref{fig:spectral2}. The lense-like quasiparticle sheets appear close to the direction of the helical momentum \bq~and grow with field strength for both Rashba sheets $\la=\pm1$.

%
%

%
%

\section{The quasiparticle interference spectrum}
\label{sec:QPI}

Now we turn to the main object of this work, the calculation of the quasiparticle interferenc spectrum in the helical phase
which should show in a very straightforward manner the effect of the sofar hypothetical momentum-space segmentation
of quasiparticles into paired and unpaired regions determined by field \bb~and Cooper pair momentum $2\bq$. This 
effect contains the microscopic essence of the helical superconducting state. For this purpose it is also necessary to
define a simple model for the surface-impurity scattering of quasiparticles and transform it to the basis of helical Rashba 
band states.
 \subsection{The normal and magnetic impurity scattering}
\label{subsec:impurity}

We consider the two most frequent cases of  normal charge (c) impurities  and  magnetic moment (m) impurities  being responsible for electron scattering at the surface of the Rashba FFLO- type superconductor.
In the normal state, using spin representation of conduction electrons the isotropic scattering  from impurities 
located at random sites $\bR_i$ is described by 
\be
\bl
&
U_c(\br-\bR_i)=U_0\sigma_0\delta(\br-\bR_i); \;\;\;
\\
&
U_m(\br-\bR_i)=U_1\sigma_z\delta(\br-\bR_i)
\el
\ee
for the charge and exchange scattering, respectively. In the latter case we assumed an Ising-type classical local moment $\bra S_z(i)\ket$ at site $i$ oriented along z-direction by a uniaxial potential, i.e. $U_1=\fs J_{\rm ex}\langle S_z(i)\rangle$ where $J_{\rm ex}$ is the on-site exchange constant.
This leads to a Hamiltonian
\be
\bl
H_{\rm imp}=\sum_{i\bk\bk'\sigma}U_\sigma a_{\bk'\sigma}^\dag a_{\bk\sigma}e^{i(\bk'-\bk)\bR_i};\;\;\;
U_\sigma=U_0+\sigma U_1
\el
\ee
in spin representation describing the scattering by random impurities at the surface where $\tbq=\bk'-\bk$ is the momentum transfer. It has to be transformed to the helical eigenstates of the Rashba bands
defined by Eq.~(\ref{eq:helical}). Furthermore in the superconducting state we must
use appropriate scattering matrices in Nambu (particle-hole)  space according to the replacement $(U_0,U_1)\rightarrow (U_0\tau_z, U_1\tau_0)$ where $\tau_z,\tau_0$ are corresponding Pauli and unit matrices, respectively \cite{akbari:13}. Then we obtain
\be
\bl
H_{\rm imp}
=
\sum_{i\bk\bk'\la\la'}
\!\!
\Big[&
U_0\tV^0_{\la\la'}(\bk\bk')\tau_z
\\
&
+U_1\tV^1_{\la\la'}(\bk\bk')\tau_0
\Big]
c^\dag_{\bk'\la} 
c_{\bk\la'}e^{i(\bk'-\bk)\bR_i}
.
\label{eq:himphel}
\el
\ee
Here the momentum dependent scattering form factors $\tV^{0,1}_{\la\la'}(\bk\bk')$ are introduced by the transformation to helical eigenstates $|\bk\la\ket$ of each Rashba bands according to Eq.~(\ref{eq:helical}). They are obtained from the transformation matrix in this equation according to
\be
\bl
&
\tV^0_{\la\la'}(\bk\bk')=
\sum_\sigma S^*_{\sigma\la}(\bk')S_{\sigma\la'}(\bk); \;\;\; 
\\&
\tV^1_{\la\la'}(\bk\bk')
=
\sum_\sigma \sigma S^*_{\sigma\la}(\bk')S_{\sigma\la'}(\bk)  .
\el
\ee
Explicitly we obtain in helicity space $(\lambda\lambda')$ effective momentum- dependent c,m scattering potentials, given, respectively by
\be
\bl
&
\{\tV^0_{\la\la'}(\bk\bk')\}
=
\fs\left[
\begin{matrix}
1+e^{i(\theta_\bk-\theta_{\bk'})}&i(e^{-i\theta_\bk}-e^{-i\theta_{\bk'}}) \cr
i(e^{i\theta_\bk}-e^{i\theta_{\bk'}}) & 1+e^{-i(\theta_\bk-\theta_{\bk'})}
 \end{matrix}\right];\;\;
 \\[0.3cm]
 &
 \{\tV^1_{\la\la'}(\bk\bk')\}=
\fs \left[
\begin{matrix}
1-e^{i(\theta_\bk-\theta_{\bk'})}&i(e^{-i\theta_\bk}+e^{-i\theta_{\bk'}}) \cr
-i(e^{i\theta_\bk}+e^{i\theta_{\bk'}}) & -(1-e^{-i(\theta_\bk-\theta_{\bk'}))}
 \end{matrix}\right].
 \label{eq:formfac}
\el
\ee
These scattering matrices are Hermitean fulfilling the relations $\tV^{\kappa*}_{\la\la'}(\bk,\bk')=\tV^\kappa_{\la'\la}(\bk',\bk)$ $(\kappa =0,1)$.
Since we use the periodic TB band model for the QPI calculation in the next section we also must use the periodic form 
of the phase angle  $\theta_\bk=\tan^{-1}(\sin k_y/\sin k_x)$ in the above expressions  appropriate for the TB model. Note that one has to be careful to pick the right branches so that the polar angle covers the whole interval $[0,2\pi]$. This is guaranteed if we define $\theta^0_\bk=\tan^{-1}(|\sin(k_y)/\sin(k_x)|)$ and choose $\theta_\bk$ in the whole BZ $-\pi\leq k_x,k_y\leq\pi$ in counterclockwise fashion in the four quadrants (I-IV) $(\pm k_x>0, \pm k_y >0)$ according to
\be
\bl
&
\mbox{(I)}: \; \theta_\bk=\theta^0_\bk;\;
\quad\quad \quad
\mbox{(II)}:\; \theta_\bk=\pi-\theta^0_\bk;\;\;\;
\\
&
\mbox{(III)} :  \; \theta_\bk=\theta^0_\bk+\pi ;\;\;\;
\mbox{(IV)} : \; \theta_\bk=2\pi-\theta^0_\bk.
\el
\ee
In the helicity representation the scattering matrix  includes non-diagonal inter-band terms $\la\neq\la'$ even though we started from a scattering potential diagonal in spin quantum numbers. Both momentum dependence and interband features of the scattering play a role in the QPI spectrum.

\subsection{QPI spectrum in Born approximation}
\label{subsect:Born}

The Fourier component of the surface charge modulation corresponding to momentum transfer $\tbq=\bk'-\bk$ (not to be confused with Cooper pair momentum $2\bq$) and bias voltage $\omega =eV$ induced by the scattering from random impurites is  given by (per impurity site)\cite{akbari:13,Akbari:2013tt}

\be
\bl
&
\delta N(\tbq,\omega)=-\frac{1}{\pi}
 {\rm Im} 
 {\Big [}
 \tL(\tbq,i\omega_n)
 {\Big]}_{i\omega_n\rightarrow \omega + i\delta},
\\
&
\tL(\tbq,i\omega_n)=
\\&\quad
\frac{1}{N}\sum_{\bk\la\la'}
{\Big [}
\tau_z\hG_\la(\bk ,i\omega_n)\hatt_{\la\la'}(\bk\bk',\om)\hG_{\la'}(\bk',i\omega_n)
{\Big ]}_{11},
\label{eq:QPI}
\el
\ee
where $\hatt_{\la\la'}(\bk\bk')(\om)$ is the scattering t- matrix due to the impurity scattering potential (Eq.~(\ref{eq:himphel})) and the index $(11)$ projects out the electron part of the Nambu matrix. Since the effective scattering in $H_{\rm imp}$  is 
momentum dependent due to helical transformation  we treat it only in Born approximation (BA) for weak scattering. As a matter of experience the QPI spectra in momentum space do not strongly depend on this simplification~\cite{akbari:13}. In the Born case the t- matrix is frequency independent and simply given by
\be
\hatt^c_{\la\la'}(\bk\bk')=U_0\tV^0_{\la\la'}(\bk\bk')\tau_z; \quad
\hatt^m_{\la\la'}(\bk\bk')=U_1\tV^1_{\la\la'}(\bk\bk')\tau_0,
\ee
in the normal (charge) and exchange (magnetic) scattering (c,m) cases, respectively. Inserting this in Eq.~(\ref{eq:QPI}), using the explicit
FFLO- type Green's function (Eq.~(\ref{eq:gmat})) and defining $\tL_{(0,1)}(\tbq,i\omega_n)=U_{(0,1)}\Lambda_{0,1}(\tbq,i\omega_n)$ we obtain the final result of QPI spectrum function (suppressing the pair momentum index \bq~ everywhere)
\begin{widetext}
\bea
\tL_\kappa(\tbq,i\omega_n)=\frac{1}{N}\sum_{\bk\la\la'}\tV^{\kappa}_{(1)\la\la'}(\bk\bk')
\Bigl[\frac{(\om+\vare^s_{\bk\la}- \vare^a_{\bk\la})(\om+\vare^s_{\bk'\la'}- \vare^a_{\bk'\la'})
-(-1)^\kappa\Delta_\la\Delta_{\la'}}
{(\om-E^+_{\bk\la})(\om+E^-_{\bk\la})(\om-E^+_{\bk'\la'})(\om+E^-_{\bk'\la'})}\Bigr]
\label{eq:Lambda}
\eea
\end{widetext}
for the two cases of  normal $(c, \kappa =0)$ and magnetic $(m, \kappa=1)$ scattering, respectively, whereby the sign constraint $\Delta_+=-\Delta_-=\Delta(q,b)$ for the gap functions has to be kept. In this sum we are using the BA scattering matrix from Eq.~(\ref{eq:formfac}) the quasiparticle energies from Eq.~(\ref{eq:SCquasi}) and the (anti-) symmetrized normal state dispersions from  Eq.~(\ref{eq:symeps}). The value of the SC gap is obtained from the minimization procedure of Eq.~(\ref{eq:condens}). Note that only the real part of the scattering matrix $\tV^{\kappa}_{(1)\la\la'}(\bk\bk')={\rm Re}\tV^{\kappa}_{\la\la'}(\bk\bk')$ which, due to the hermiticity of  Eq.~(\ref{eq:formfac}),  is symmetric under exchange of all indices enters the expression for $\tL_\kappa(\tbq,i\omega_n)$. Likewise  the imaginary part  $\tV^{\kappa}_{(2)\la\la'}(\bk\bk')={\rm Im}\tV^{\kappa}_{\la\la'}(\bk\bk')$ must be antisymmetric under this  exchange and beacuse the expression in parentheses in Eq.~(\ref{eq:Lambda}) is symmetric the summation over it gives zero. The real symmetric scattering matrix elements in Eq.~(\ref{eq:Lambda}) in the charge $(\kappa=0)$ and magnetic $(\kappa=1)$ impurity cases are obtained from Eq.~(\ref{eq:formfac}) as
\be
\bl
&
\{\tV^0_{(1)\la\la'}(\bk\bk')\}
\!=\!
\!
\fs\left[
\begin{matrix}
1+\cos(\theta_\bk-\theta_{\bk'})
&
\sin\theta_\bk-\sin\theta_{\bk'} \cr
-\sin\theta_\bk+\sin\theta_{\bk'}
&
 1\!+\!\cos(\theta_\bk-\theta_{\bk'})
 \end{matrix}\right]
 ;
\\[0.3cm]
&
 \{\tV^1_{(1)\la\la'}(\bk\bk')\}
 \!=\!
\!
\fs
\!
\left[
\begin{matrix}
1
\!-\!
\cos(\theta_\bk-\theta_{\bk'})
&
\sin\theta_\bk+\sin\theta_{\bk'} \cr
\sin\theta_\bk+\sin\theta_{\bk'} 
&
 -1\!+\!\cos(\theta_\bk-\theta_{\bk'})
 \end{matrix}\right].\\
 \label{eq:formfacreal}
\el
\ee
%
The difference between the two is due to the different influence of helical spin texture in the two scattering mechanisms.

\section{Discussion of numerical QPI results: the STM image of momentum space segmentation}
\label{sec:discussion}

%
\begin{figure}
\includegraphics[width=0.990\linewidth]{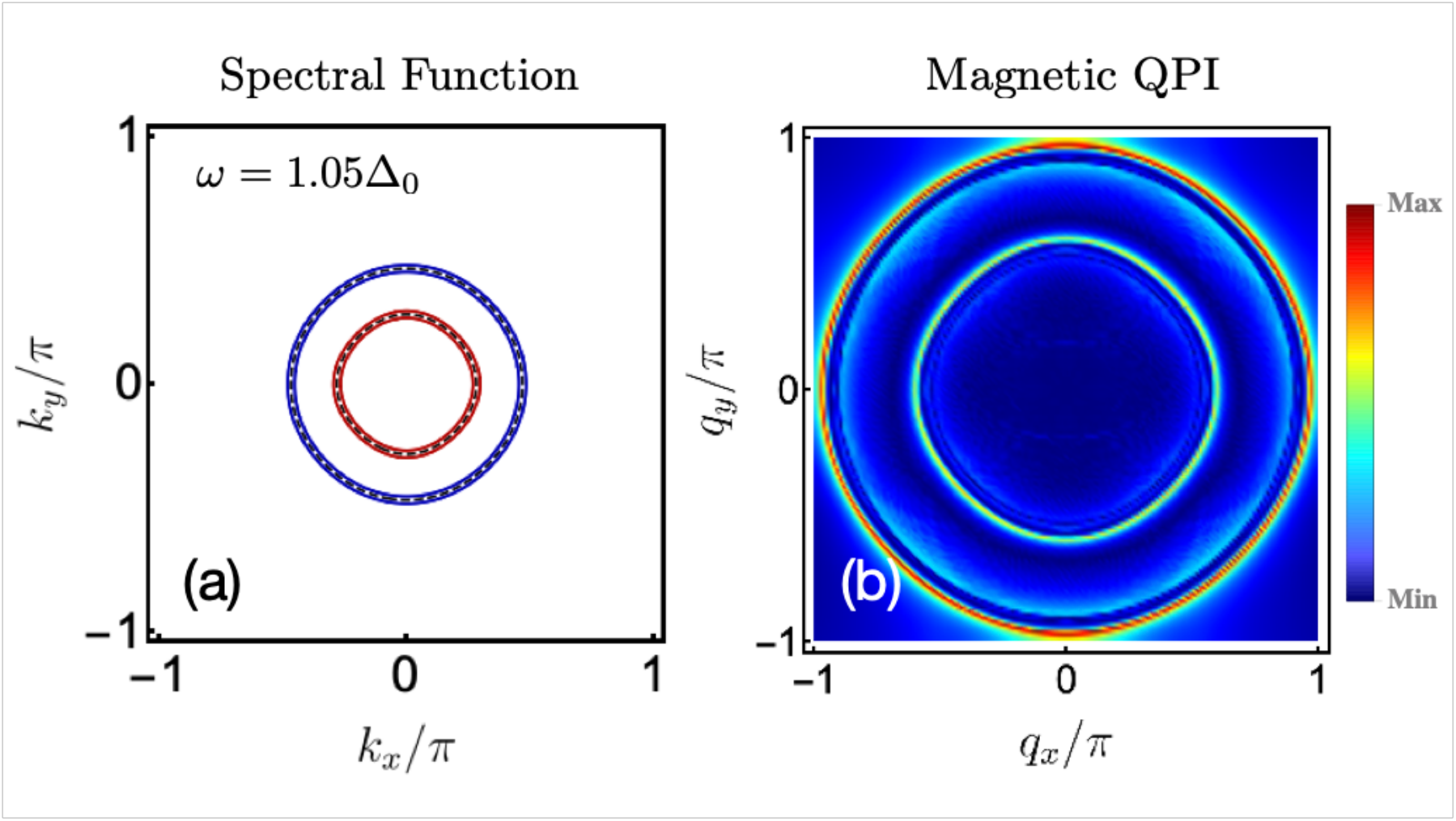}
\caption{
Reference spectral function (a) and  QPI spectrum (b) for Rashba-BCS phase
 $(b,q=0;\; \Delta_0=0.3t,\;\alpha=2\Delta_0)$ 
for $\omega\simeq \Delta_0$.
}
\label{fig:QPI-BCS}
\end{figure}
%

The QPI method is well suited to observe the typical changes of qausiparticle sheets in momentum space connected with the appearance of the FFLO-type helical phase. The most characteristic feature is the reappearance of Fermi surface sheets for small frequencies $\omega < \Delta_0$ due to the pair breaking of combined Zeeman shift and Rashba splitting effects. The latter happens primarily for Cooper pairs with momenta $-\bk+\bq, \bk+\bq$ close to the direction of the shift vector $\bq_s$ of Rashba Fermi surfaces which is perpendicular to the applied field. As function of applied QPI voltage $\omega=eV$ and field strength the unpaired sheets represented by the quasiparticle spectral functions undergo typical changes which contain information about the microscopic structure of the helical state. In particular it will give direct evidence for the Cooper pair momentum $2\bq$ being perpendicular to the applied field and more importantly under favorable conditions it should be possible to estimate its magnitude from analysing characteristic momenta $\tbq_i$ of the QPI image.\\

In the following we will therefore discuss the typical QPI charge images $\delta N(\tbq,\omega)$ expected in experiment which we derived in the previous section for the charge and magnetic impurity scattering cases. It will turn out that the two are to a certain extent complementary. They will present mainly the same features due to the same quasiparticle energy denominators in Eq.~(\ref{eq:Lambda}) but with different intra-/inter- band  intensity distribution due to the  coherence factors in the numerator which contain different signs  $(-1)^\kappa$ for the two scattering mechanisms.  Furthermore the momentum dependence of  effective scattering matrices in  Eq.~(\ref{eq:formfacreal}) is different in the two cases. In order to achieve sufficient numerical accuracy for 
detailed QPI image structure we have to use an enhanced size for the SC gap scale $\Delta_0$ which will be set
to $0.5\alpha$ throughout.\\

Firstly, as a reference, we will briefly discuss the QPI image in the zero-field BCS case with conventional Cooper pairs, i.e. $\bq=0$, of Fig.~\ref{fig:QPI-BCS} (see also Ref. \cite{akbari:16}). In (a) the spectral function presents two almost isotropic and featureless Rashba-split Bogoliubov quasiparticle sheets (full lines) which show an additional splitting due to the doubling of particle-hole branches by the superconducting gap. For frequencies $\omega$ slightly above the gap size their radii are close  to the Fermi wave vectors $k_F^\lambda$ of the normal state (dashed lines) given in Sec.~\ref{subsect:normalmodel}. In this case it is well known that the QPI image generated by all scattering events accross the two spheres is again spherical with the doubling of the radius to approximately $2k^\lambda_F$ as is indeed seen in Fig.~\ref{fig:QPI-BCS}. For frequencies $\omega$ slightly below  the gap $\Delta_0$ this QPI image is rapidly 
extinguished.\\ 

\begin{figure}
\includegraphics[width=\linewidth]{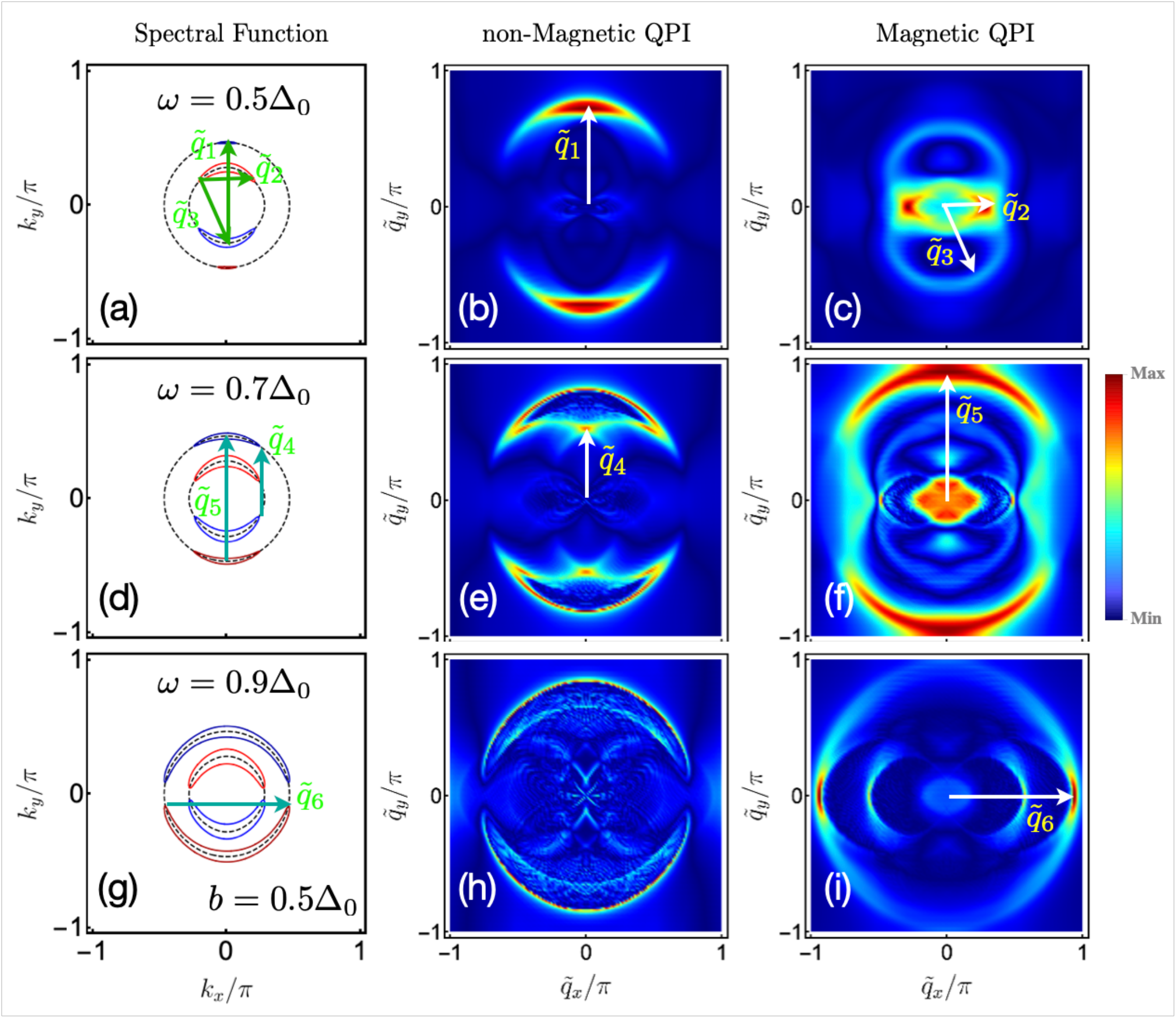}
\caption{
Evolution of spectral function (a, d, and g)
and 
corresponding QPI spectrum for Rashba-helical FFLO phase:
(b,e, and h) charge scattering QPI, (c,f, and i) magnetic scattering QPI,
with frequency [first row: $\omega=0.5\Delta_0$; second row: $\omega=0.7\Delta_0$; third row: $\omega=0.9\Delta_0$],
 and  at field $b=0.5\Delta_0$ with  $q(b)/\pi=0.005 $, and $\Delta_q(b) = \Delta_0$.
Note that inter-band scatterings are contributed mostly from the non-magnetic impurities, whereas magnetic impurities  mainly lead to  intra-band scatterings.
}
\label{fig:QPI-1}
\end{figure}

\begin{figure}
\includegraphics[width=\linewidth]{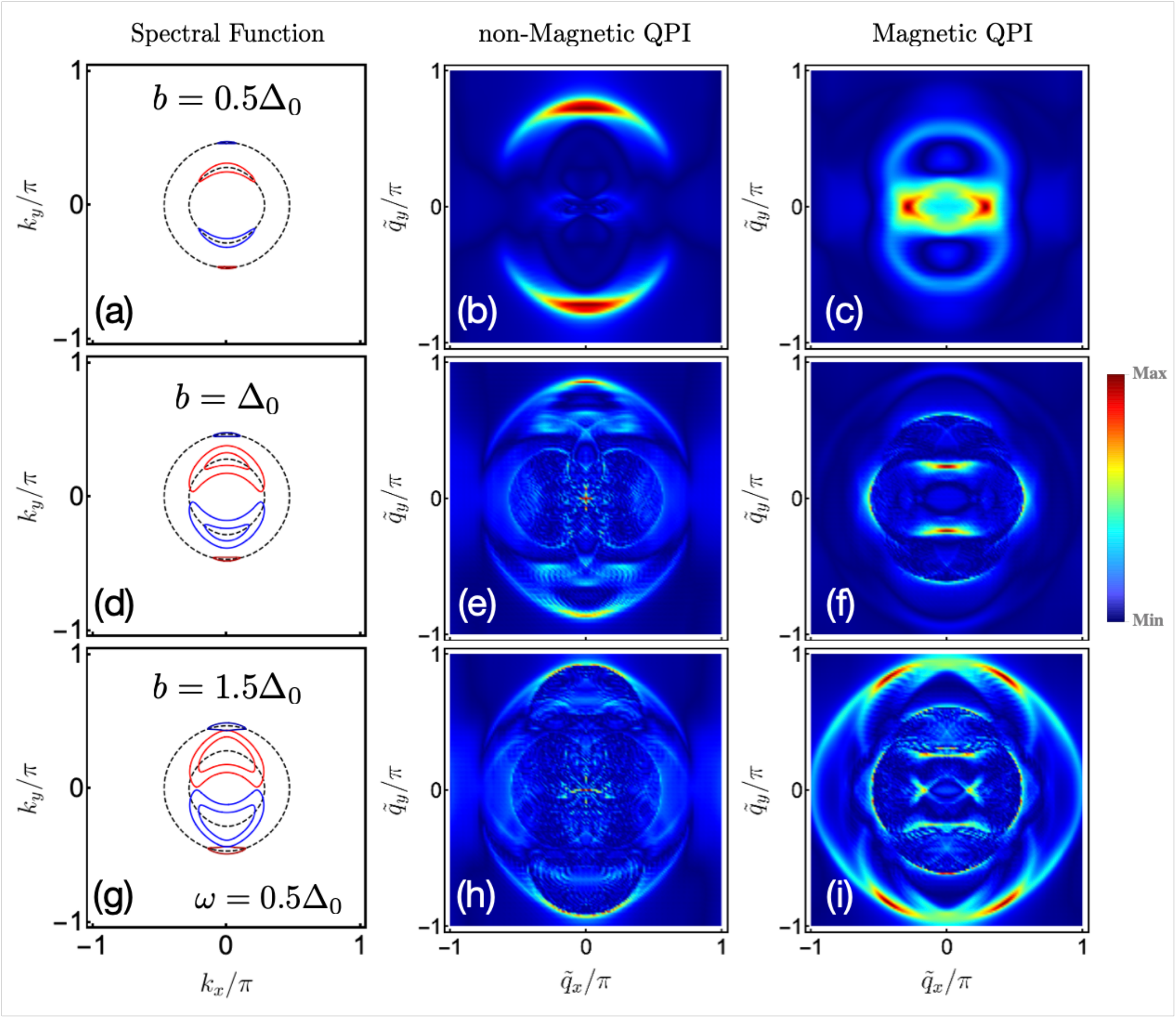}
\caption{
Evolution of spectral function (a, d, and g)
and 
Corresponding QPI spectrums for Rashba-helical FFLO phase:
(b,e, and h) charge scattering QPI, (c,f, and i) magnetic scattering QPI,
with field b and at frequency $\omega=0.5\Delta_0$.
The first, second, and third  rows correspond to
$b=0.5\Delta_0$ with $q(b)/\pi=0.005 $, and $\Delta_q(b) = \Delta_0$;
 $b=\Delta_0$ with $q(b)/\pi=0.035 $, and $\Delta_q (b)= 0.75\Delta_0$;
 and
$b=1.5\Delta_0$ with $q(b)/\pi=0.06 $, and $\Delta_q(b) = 0.59\Delta_0$,
respectively.
%
%
%
%
}
\label{fig:QPI-2}
\end{figure}

In distinction in the helical phase with superconducing order parameter $\Delta_{\bq\la}$ corresponding to finite pair momentum $2\bq$ the regions in $\bk$ space where  Bogoliubov energy $E^+_{\bk\bq\la}<0$ or  $E^-_{\bk\bq\la}<0$ are depaired and have normal quasiparticle energies  $|E^+_{\bk\bq\la}|$ or  $|E^-_{\bk\bq\la}|$ starting from zero and hence lead to quasiparticle sheets even for $\omega<|\Delta_{\bq\la}|$. They are presented by plotting the spectral functions of Eq.~(\ref{eq:specfunc}) for various bias voltage $eV=\omega$ or frequencies in the left columns of Figs.~\ref{fig:QPI-1}, \ref{fig:QPI-2} (see also Fig.~\ref{fig:spectral2}). The segmentation of \bk- space into paired regions without low-energy quasiparticles (small $|k_y|$) and unpaired regions with quasiparticle sheets (large $|k_y|$) is clearly seen for the different frequencies. Here the inner/outer Rashba FS (dashed lines) correspond to $\la=\pm1$ and the blue/red bent lenses to quasiparticle sheets $|E^\tau_{\bk\bq\la}|=\omega$ correspond to  $\tau=\pm 1$. For small $\omega$ (a) the first sheet appears in the inner Rashba band $\la=-1$ and then on the outer one $\lambda=+1$ increasing in size with increasing $\omega$ (d,g). They are ending at the tip positions characterized by polar angles $\theta_\bk$ where $|E^\tau_{\bk\bq\la}|=\omega$ and $|\bk|\simeq k_F^\la$. The large curvature at these points leads to a small group velocity and hence large DOS contribution from their vicinity. Hence they may appear prominently  in the integrated QPI spectrum, however as mentioned before the momentum dependent scattering matrix elements also influence the intensity.\\ 

 %
%
%
%

From a comparison of the model calculation for the segmented Fermi surfaces (more precisely equal-energy surfaces at bias voltage $\omega=eV$) and its associated theoretical QPI spectrum with the experimental one it is possible to investigate
the details of the pair-breaking effect in the helical phase on the quasiparticle spectrum.
In Fig.~\ref{fig:QPI-1} we give a comparison between calculated spectral functions left column) in the helical phase and its associated predicted QPI spectra and show their evolution  as function of frequency or bias voltage for constant field (center and right column for charge and magnetic impurity scattering, respectively).
We can identify a selection of the characteristic possible intra- $(\la=\la')$ and inter- $(\la\neq\la')$ band scattering 
vectors $\tbq_i$ $(i=1-6)$ defined in the left column as intense or at least enhanced features in the QPI image in the center and right column. These 
correspondences are indicated in the panels with white arrows. Particularly prominent and easy to identify are the tip-to-tip scattering vectors $\tbq_4$ for nonmagnetic and $\tbq_2,\tbq_6$ for magnetic scattering. The other characteristic QPI vectors map out whole Fermi surface arc segments of the spectral function in the left column. In reverse this means that an experimental QPI spectrum in 
the helical phase of a Rashba superconductor allows one to reconstruct the segmented Fermi surface sheets that appear as a consequence of the depairing of Cooper pairs whose momenta are primarily oriented along the helical \bq~vector.
It is also noteworthy that  the intensity distribution of the QPI spectrum is to a certain extent complementary for non-magnetic and  magnetic scattering, emphasizing different regions of $\tbq$- space: 
The inter-band scatterings appear most prominent for non-magnetic impurities, whereas magnetic impurities  mainly lead to  intra-band scatterings. This is due to the different coherence factors (numerators) in Eq.~(\ref{eq:Lambda}) and angular dependences of the effective  scattering matrices in Eq.~(\ref{eq:formfacreal}) for the two cases.\\

The Fig.~\ref{fig:QPI-2} presents  results for the field evolution of QPI as an alternative to the previous one. Now the frequency is fixed to $\omega=0.5\Delta_0$ and the field is varied in the low field regime $b < \alpha$ of the helical phase (the zero field BCS case is already presented in Fig.~\ref{fig:QPI-BCS} and the first row is identical  to the one in  Fig.~\ref{fig:QPI-2}). Whereas in the previous figure the quasiparticle sheets
simply extend their dimension along the Rashba circle with increasing $\omega$ now the increasing field changes their shape and may lead to a doubling. This means the field evolution of the QPI spectrum in central and right column are also distinct. It is again possible to identify characteristc scattering vectors in the latter that correspond to those connecting the various sheets in the spectral function.\\

Altogether our analysis demonstrates that an experimental magnetic/nonmagnetic QPI spectrum and its frequency and field evolution should contain enough information to map out the segmented quasiparticle sheets in the helical phase with finite Cooper pair momentum which is at the heart of this FFLO-type Rashba superconducting state.\\

Finally one may ask whether the information contained in the QPI images allows to extract the size of the Cooper pair momentum \bq~as function of field from the experimental data . We note that  none of the thermodynamic experimental methods can achieve this. Since the \bq-vector for moderate fields has only a small fraction of the
BZ extension and because it enters in a complicated manner in the  spectrum of Eq.~(\ref{eq:Lambda}) one may not expect a direct identification in the QPI images of Figs.~\ref{fig:QPI-1},\ref{fig:QPI-2}. However it is possible to derive an empirical relation for its estimation from experimental quantities for small fields. For this purpose we note that the frequency dependent
tips of the spectral functions at polar angles $\theta^\tau_{\bk \la}$   in the left column of Figs.~\ref{fig:QPI-1}, \ref{fig:QPI-2} are characterized by the following conditions i) their quasiparticle energy fulfils $|E^\tau_{\bk\bq\la}|=\omega$ and ii) they lie very close to the original (dashed lines) Rashba Fermi spheres with radius $k_F^\la$ (Sec.~\ref{subsect:normalmodel}). On these spheres Eq.~(\ref{eq:qpera2}) reduces to
\be
E^\tau_{\bk\bq\la}\simeq \Delta_\bq+\tau(v_Fq+\la b)\sin\theta^\tau_{\bk\la}\equiv\omega.
\label{eq:empiricalq}
\ee
We can determine the angles $\theta_{\bk\la}^\tau$ at the tip positions  from the geometry depicted in Fig.~\ref{fig:QPI-1}(a,d,g). To be specific let us consider the upper part $(\tau=-1)$ of the inner sheet $(\la=-1)$ extended along the Rashba sphere with radius $k^-_F$. Its right $(\theta_R)$ and left  $(\theta_L=\pi-\theta_R)$ tips are connected by characteristic vector $\tbq_2$ which is prominently seen in corresponding magnetic QPI spectrum  (Fig.~\ref{fig:QPI-1}(c)). Then we obtain $\cos\theta_R=\frac{\tilde{q}_2}{2k_F^-}$. The sheet with a value $0<\theta_R<\pi/2$ exists only when $\omega > \Delta_q-(v_Fq-b) \equiv \omega_0$ or equivalently when $\omega'=\omega-\omega_0 >0$. Then we may resolve Eq.~(\ref{eq:empiricalq}) to obtain a phenomenological 
\be
q(b,\omega')=\frac{b}{v_F}+\frac{\omega'}{1-\sin\theta_R(\omega')},
\ee
where the first term is the Rashba FS shift $q_s$ of Eq.~(\ref{eq:Rshift}). The Cooper pair momentum $q(b)$  is then obtained from the extrapolation to small $\omega'\rightarrow 0 $ where  $\sin\theta_R(\omega')=[1-\frac{\tilde{q}_2^\la}{2k_F^\la}]^\fs\rightarrow 1$
in this limit. It has to be obtained from the experimentally observed $q_2(\omega')$.
A similar  procedure may be applied to other characteristic QPI vectors $\tbq_i$ to obtain q(b). In principle this opens a way to determine the Cooper pair momentum $2q(b)$ directly from STM-QPI experiments.

\section{Conclusion and Outlook}
\label{sec:conclusion}

In this work we investigated microscopic features of helical phase in Rashba superconductors with isotropic and equal magnitude of the gap function on the two Rashba bands. The latter have helical spin texture enforced by the strong Rashba spin-orbit coupling. In a magnetic field they are shifted perpendicular to the field by an amount proportional to its size.
Therefore Cooper pairing in a state with non-vanishing pair momentum 2\bq~will be favored.

Using the approximations for large Rashba coupling we derived the condensation energy as function of \bq. Minimization leads  to the dependence of pair momentum and gap size on the applied field. At the same time we computed the quasiparticle energies in the helical state. Their most interesting aspect is a segmentation of momentum space into regions where Cooper pairs are stable and gapped Bogoliubov excitations exist and other regions spread around the direction of the overall pair momentum where pair breaking due to large kinetic energy destroys the Cooper pairs and leads to normal low- energy quasiparticles with corresponding Fermi surface sheets. These are present despite the fact that the gap $\Delta^\bk_{\bq\la}=\Delta_{\bq\la}$ is nodeless in \bk- space and real space.

This basic microscopic structure of the helical state, a coherent superposition of paired and unpaired states with associated peculiar evolution of Fermi surface  (surfaces of constant energy) topology as function of field and frequency has sofar not been investigated experimentally. In this work we have shown that the technique of quasiparticle interference is well suited to address this central property of Rashba superconductors with finite momentum Cooper pairing. It  is able to monitor the apperance of the segmented Fermi surface sheets of unpaired quasiparticles as function of field strength and bias voltage until they evolve
into those of the normal state Rashba sheets for large values of these tuning parameters. Due to the helical frozen spin texture the QPI images obtained for charge and magnetic impurity scattering on the surface show considerable difference and are complementary in the intensity distribution.
Furthermore following some of the characteristic wave vectors of the segments one may derive an estimate for the size of
the pair momentum 2q which is not accessible by other experimental means.

The FFLO-type helical phase  in the Rashba superconductor is more amenable to such QPI investigations because 
it appears already for small fields and does not require the extremely large fields of the genuine FFLO phase in the
inversion symmetric superonductors. It may also occur more frequently since there is a considerable number of inversion-symmetry breaking (non-centrosymmetric) superconductors known by now. In particular such QPI investigations for the 
helical phase should be possible in layered superonductors with strong 2D character which has been assumed in our analysis.

\section*{Acknowledgments}
A.A. acknowledges the support of the Max Planck POSTECH/Hsinchu Center for Complex Phase Materials. 


\appendix

\section{Derivation of the superconducting condensation energy}
\label{sec:GSenergy}

Here we give a brief derivation of Eq.~(\ref{eq:condens}) used to find the $(\bq,\Delta_\bq)$ values by minimization.
First we note that the ground state energy $\langle H_{\rm BCS}\rangle$ for the paired states (first row in curly brackets in Eq.~(\ref{eq:bcsbogol})) may also be written in different equivalent forms given below:
\be
\bl
\langle H_{\rm BCS}\rangle_{\rm paired}
=& \frac{1}{2}\sum_{\bk\la}{'}
[\vare^s_{\bk\bq\la}-E_{\bk\bq\la}+\frac{|\Delta_{\bq\la}|^2}{V_0}] 
\\=&
\frac{1}{2}\sum_{\bk\la}{'}
[2\vare^s_{\bk\bq\la}v^2_{\bk\la}-\frac{|\Delta_{\bq\la}|^2}{V_0}]
\\
=&\frac{1}{2}\sum_{\bk\la}{'}[\vare^s_{\bk\bq\la}-E_{\bk\bq\la}+\frac{|\Delta_{\bq\la}|^2}{2E_{\bk\bq\la}}] \
\\=&
\frac{1}{2}\sum_{\bk\la}{'}
[\vare^s_{\bk\bq\la}
-\frac{\vare^{s2}_{\bk\bq\la}}{E_{\bk\bq\la}}
-\frac{|\Delta_{\bq\la}|^2}{2E_{\bk\bq\la}}] .
\label{eq:enequiv}
\el
\ee
Here the prime denotes summation over paired states only with both $E_{\bk\bq\la}^\pm >0$.
Using the first form above the total ground state energy obtained from the mean field approximation and Bogoliubov transformation is orginally given by
\be
\bl
&E_G(\bq,\Delta_{\bq\pm})=
\fs\sum_\la\Bigl[
N\bigl(\frac{|\Delta_{\bq\la}|^2}{V_0}\bigr)
+\sum_\bk\vare^s_{\bk\bq\la}
\\&+
2\sum_\bk\vare^a_{\bk\bq\la}\Theta(-E^+_{\bk\bq\la})
-\sum_\bk E^-_{\bk\bq\la}\Theta(E^+_{\bk\bq\la})\Theta(E^-_{\bk\bq\la})\Bigr].
\label{eq:GSenas}
\el
\ee
In the zero-field normal state $(b=0,\bq=0, \Delta_{\bq\la}=0)$ where $\vare^a_{\bk\bq\la}=0$ and $E^\pm_{\bk\bq\la}=|\vare^0_{\bk\la}| >0$ this ground state energy reduces to
\bea
\bl
&
E^0_G=\fs\sum_{\bk\la}(\vare^0_{\bk\la}-|\vare^0_{\bk\la}|)=\sum_{\bk\la}f_{\bk\la}\vare^0_{\bk\la},
\\&
\vare^0_{\bk\la}=\vare^s_{\bk\bq\la}(\bq=0,b=0)=\xi_\bk+\la|\alpha\bg_\bk|,
\label{eq:gsenor2}
\el
\eea
where $f_\bk=\Theta(-\vare^0_{\bk\la})$ is the zero temperature Fermi function for the unpolarized Rashba-split bands $\vare^0_{\bk\la}$ (cf. Eq.~(\ref{eq:Rdisp0})). The condensation energy for the minimization is then given by $E_c=E_G-E_G^0$.\\

To obtain a more symmetric form for $E_G$ and $E_c$ we now use the identity
\be
\Theta(E^+_{\bk\bq\la})\Theta(E^-_{\bk\bq\la})
=1-\Theta(-E^+_{\bk\bq\la})-\Theta(-E^+_{\bk\bq\la}),
\label{eqn:thetaid}
\ee
which holds because both $E^\pm_{\bk\bq\la}$ cannot be simultaneously negative since their sum $E^+_{\bk\bq\la}+E^-_{\bk\bq\la}=E_{\bk\bq\la} >0$. Inserting this into Eq.~(\ref{eq:GSenas}) and using $\sum_\bk\vare_{\bk\bq\la}^a=0$ we obtain after some simple rearrangements  the  symmetrized form of the ground state energy
\be
\bl
E_G(\bq,\Delta_{\bq\pm})=
&
\fs\sum_\la\Bigl[
N\bigl(\frac{|\Delta_{\bq\la}|^2}{V_0}\bigr)
-\sum_\bk(E_{\bk\bq\la}-\vare^s_{\bk\bq\la})
\\&
+\sum_\bk[E^+_{\bk\bq\la}\Theta(-E^+_{\bk\bq\la})+E^-_{\bk\bq\la}\Theta(-E^-_{\bk\bq\la})]\Bigr]
\label{eq:GSen2}
\el
\ee
given before in Eq.~(\ref{eq:GSen}). Subtracting the normal state energy of Eq.~(\ref{eq:gsenor2}) we obtain again the condensation energy expression given in Eq.~(\ref{eq:condens}).
\\

\section{The proof of vanishing charge current}
\label{sec:chargecurrent}

Without a Rashba coupling it is known that the total charge current in the helical state vanishes even though the pairs have finite momentum. This is due to the fact that the current is the pair-momentum derivative of the total energy which must vanish in the ground state~\cite{cui:06}. Here we show that this still holds for the case of finite Rashba coupling. The charge current operator is commonly given in terms of Bloch operators creating spin $\sigma_z$ eigenstates~\cite{cui:06}. After a unitary transformation to helical states $(\la=\pm$) in the Rashba system we obtain (in units of e):
\be
{\bf J}^c_\bq
=
\!
\frac{1}{m}\sum_\bk
[(\bk+\bq)c^\dg_{\bk+\bq +}c_{\bk+\bq+} 
- (\bk-\bq)c^\dg_{-\bk+\bq -}c_{-\bk+\bq-} 
].
\label{eq:current1}
\ee
Transforming to Bogliubov quasiparticle states with Eqs.~(\ref{eq:BGLp},\ref{eq:BGLu}) we obtain for the y-component ($q_y=q, k_y=k$) of the current;
\be
\bl
\langle J^c_q\rangle
=&
\frac{1}{2m} \sum_{\bk\la}
\Big[
2q|v_{\bk\la}|^2\theta_H(E^+_{\bk\bq\la})\theta_H(E^-_{\bk\bq\la})
\\&+
(q+k)\theta_H(-E^+_{\bk\bq\la})+
(q-k)\theta_H(-E^-_{\bk\bq\la})
\Big].\\
\label{eq:current2}
\el
\ee
Now we consider again the total ground state energy Eq.~(\ref{eq:bcsbogol}), using an equivalent form for the paired term
according to Eq.~(\ref{eq:enequiv}) and the relation  $\sum_{\bk\la}\vare^a_{\bk\bq\la}=0$:
\be
\langle H_{\rm BCS}\rangle=
\frac{1}{2}\sum_{\bk\la}
\left\{
 \begin{array}{c}
2\vare^s_{\bk\bq\la}v^2_{\bk\la}-\frac{|\Delta_{\bq\la}|^2}{V_0};
\;\;\;\;\;\;\;
E^\tau_{\bk\bq\la}>0\\
\vare_{\bk\bq\la}(b)+\frac{|\Delta_{\bq\la}|^2}{V_0};
\;\;\;\;\; \;
\quad
E^+_{\bk\bq\la}<0\\
\vare_{\bk-\bq\la}(-b)+\frac{|\Delta_{\bq\la}|^2}{V_0};
\;\;\;\;\;
E^-_{\bk\bq\la}<0
\end{array}
\right\}.
\\
\ee
Then, using similar small-q approximation as in Sec.~\ref{sec:Green}, we arrive at the identity
\be
\frac{\partial\langle H_{\rm BCS}\rangle}{\partial q}=
\frac{1}{2}\sum_{\bk\la}
\left\{
 \begin{array}{c}
2qv^2_{\bk\la};
\;\;\;
E^\tau_{\bk\bq\la}>0\\
q+k;
\;\;\; E^+_{\bk\bq\la}<0\\
q-k;
\;\;\;
E^-_{\bk\bq\la}<0
\end{array}
\right\}=\langle J^c_{q}\rangle
.
\ee
Therefore, similar as in the inversion symmetric case the charge current is the momentum gradient of the total energy, cf. Eq.(\ref{eq:current2}). We conclude that also in the presence of the Rashba coupling we have vanishing charge current $\langle J^c_{q}\rangle=0$ in the ground state. This situation may be different for the spin current  which is already nonzero in the zero field phase
of the Rashba superconductor \cite{vorontsov:08}.

\bibliography{References}

\end{document}